\DocumentMetadata{}
\documentclass[sigconf]{acmart}

\usepackage{acmart-taps}
\usepackage{microtype}                 
\PassOptionsToPackage{warn}{textcomp}  
\usepackage{textcomp}                  
\usepackage{newtxtext,newtxmath}       
\usepackage{tabu}                      
\usepackage{booktabs}                  
\usepackage{listings}
\usepackage{enumitem}
\usepackage{multirow}
\usepackage{tabularx}
\usepackage{xcolor}
\usepackage{xspace}
\usepackage{amsmath}
\usepackage{color}
\usepackage{soul}
\usepackage{graphicx}
\usepackage{caption}
\usepackage{mathtools} 
\usepackage{fontawesome}
\usepackage{colortbl}
\usepackage{tikz}

\definecolor{orange}{RGB}{237,125,49}
\definecolor{custompurple}{RGB}{104,50,154}
\definecolor{pink}{RGB}{245,114,255}
\definecolor{black}{RGB}{0,0,0}
\definecolor{TEAL}{RGB}{0,128,128}
\definecolor{BLUE}{RGB}{0,0,255}
\definecolor{lighterdarkishgray}{RGB}{227,227,227}
\definecolor{lightergray}{RGB}{235,235,235}
\definecolor{cblue}{RGB}{0,112,192}

\definecolor{colorai}{RGB}{155,89,182}
\definecolor{colorcontrol}{RGB}{88, 88, 88}
\definecolor{colorexpert}{RGB}{39,60,117}
\definecolor{colorgroup}{RGB}{52,152,219}
\definecolor{colorguidance}{RGB}{46,204,113}

\definecolor{clightgray}{RGB}{230, 230, 230}

\newcommand{\ai}[1]{\textcolor{colorai}{\textbf{AI}}\xspace}
\newcommand{\expert}[1]{\textcolor{colorexpert}{\textbf{Expert}}\xspace}
\newcommand{\control}[1]{\textcolor{colorcontrol}{\textbf{Control}}\xspace}
\newcommand{\guidance}[1]{\textcolor{colorguidance}{\textbf{Unattributed}}\xspace}
\newcommand{\group}[1]{\textcolor{colorgroup}{\textbf{Group}}\xspace}

\newcommand{\add}[1]{#1}
\newcommand{\edit}[1]{#1}
\newcommand{\remove}[1]{}

\newcommand{\paragraphHeadingSpace}{\vspace{4px}}
\newcommand{\bpstart}[1]{
\noindent{\textbf{#1}}%
}

\newcommand{\circlednumber}[2][orange]{%
  \tikz[baseline=(char.base)]{
    \node[shape=circle,draw,#1,inner sep=0pt, minimum size=11pt] (char) {#2};}}

\AtBeginDocument{%
  }

\copyrightyear{2025}
\acmYear{2025}
\setcopyright{cc}
\setcctype{by}
\acmConference[IUI '25]{30th International Conference on Intelligent User Interfaces}{March 24--27, 2025}{Cagliari, Italy}
\acmBooktitle{30th International Conference on Intelligent User Interfaces (IUI '25), March 24--27, 2025, Cagliari, Italy}\acmDOI{10.1145/3708359.3712166}
\acmISBN{979-8-4007-1306-4/25/03}

\begin{document}

\title{Guidance Source Matters: How Guidance from AI, Expert, or a Group of Analysts Impacts Visual Data Preparation and Analysis}

\author{Arpit Narechania}
\affiliation{%
  \institution{Georgia Institute of Technology}
  \city{Atlanta}
  \state{Georgia}
  \country{USA}
}
\email{arpitnarechania@gatech.edu}
\orcid{0000-0001-6980-3686}

\author{Alex Endert}
\affiliation{%
  \institution{Georgia Institute of Technology}
  \city{Atlanta}
  \state{Georgia}
  \country{USA}
}
\email{endert@gatech.edu}

\author{Atanu R Sinha}
\affiliation{%
  \institution{Adobe Research}
  \city{Bengaluru}
  \state{Karnataka}
  \country{India}
}
\email{atr@adobe.com}

\renewcommand{\shortauthors}{Narechania, Endert, and Sinha}
\renewcommand{\shorttitle}{Guidance Source Matters}

\begin{abstract}
The progress in generative Artificial Intelligence (AI) has fueled AI-powered tools like co-pilots and assistants to provision better guidance, particularly during data analysis. However, research on guidance has not yet examined the perceived efficacy of the source from which guidance is offered and the impact of this source on the user's perception and usage of guidance. We ask whether users perceive all guidance sources as equal, with particular interest in three sources: (i) ``AI,'' (ii) ``human expert,'' and (iii) ``a group of human analysts.'' As a benchmark, we consider a fourth source, (iv) ``unattributed guidance,'' where guidance is provided without attribution to any source, enabling isolation of and comparison with the effects of source-specific guidance. We design a five-condition between-subjects study, with one condition for each of the four guidance sources and an additional (v) ``no-guidance'' condition, which serves as a baseline to evaluate the influence of any kind of guidance. We situate our study in a custom data preparation and analysis tool wherein we task users to select relevant attributes from an unfamiliar dataset to inform a business report. Depending on the assigned condition, users can request guidance, which the system then provides in the form of attribute suggestions. To ensure internal validity, we control for the quality of guidance across source-conditions. Through several metrics of usage and perception, we statistically test five \emph{preregistered} hypotheses and report on additional analysis. We find that the source of guidance matters to users, but not in a manner that matches received wisdom. For instance, users utilize guidance differently at various stages of analysis, including expressing varying levels of regret, despite receiving guidance of similar quality. Notably, users in the AI condition reported both higher post-task benefit and regret. These findings strongly indicate the need to further understand how different guidance sources impact user behavior for designing effective guidance systems.
\end{abstract}




\maketitle

\section{Introduction}
\label{section:introduction}

The prospect of offering guidance to users of a system has received fillip over the past two years as talks of Artificial Intelligence (AI)-powered tools like co-pilots and assistants caught imaginations of users and businesses alike. 
Hitherto, the guidance literature has paid attention to the important dimensions of ``why,'' ``how,'' ``what,'' and ``when'' in guided interactions~\cite{engels1996planning,perez2022typology}--contributing theories, models, frameworks, and tools on what guidance to provide, why, how, and when.
In this work, we investigate another dimension--``from whom''--focusing on the source of guidance--such as humans or an AI--which remains a gap in research.
Today, guidance is already being sought from a variety of sources across various domains.
For instance, in the analytics domain, guidance is sought from human experts (e.g., an expert analyst or consultant) or groups of peers (e.g., via community forums such as Stack Overflow\footnote{\url{https://stackoverflow.com}}, 
HuggingFace\footnote{\url{https://huggingface.co}}, 
GitHub\footnote{\url{https://github.com}}).
Relevant research on advice-taking has already studied and compared such guidance from human experts and peers~\cite{yaniv2004receiving,lai2014expert,lapple2019learning,meservy2021searching,meshi2012expert,mangin2011peer}.
Recently, there has been a growing expectation for guidance to come from an AI~\cite{vodrahalli2022humans,chong2022human,logg2019algorithm}, even though this guidance must itself rely on data from human experts or groups of peers, or other systems, to train the models and align subsequent recommendations with user preferences.
Thus, studying this ``from~whom'' dimension is important because the effect of \textit{source-attribution} in providing guidance carries significant implications for offering guidance systems.

In response, in this work, we focus on how users' perception and utility of guidance coming from a particular source impacts their performance during attribute selection as part of a visual data preparation task. In particular, we study three guidance sources--\ai{}, human \expert{}, and a \group{} of human analysts--to answer our research question, \textit{``Does the source of guidance matter to users even when the quality of guidance is held constant?''} 
As a benchmark, we also consider a fourth source, \guidance{}, wherein guidance is provided without attribution to any source, enabling isolation of and comparison with the effects of source-specific guidance.
We design a five-condition between-subjects study, with one condition for each of the four guidance sources and an additional no-guidance condition, \control{}, which serves as a baseline to evaluate the influence of any kind of guidance.
We \textit{preregister} five hypotheses of which three hypotheses compare the source-agnostic \guidance{} guidance with \control{} (no-guidance), giving a direct effect of guidance, unconfounded by the effect of any specific source. The remaining two hypotheses are source-specific, comparing guidance from \ai{} with \expert{} and \group{}, since, as argued above, these are natural candidates from whom guidance can be provisioned. Moreover, comparing each attributed source with \guidance{} isolates the impact of the attributed source.

To validate these hypotheses, we built a custom data preparation and analysis tool (as our study prototype) and tasked users to select relevant attributes from an unfamiliar dataset for a business report. Depending on the assigned condition, users can optionally request guidance, one by one, up to a maximum of ten times, which the system then provides in the form of attribute suggestions (hereafter, \emph{guided~attributes}). 
These guided attributes are a subset of all attributes in the dataset that the user can select from.
To ensure internal validity, we control for the quality of guidance by always provisioning seven relevant attributes and three irrelevant attributes, predetermined by the study team, randomly picked and sequenced, while varying the source, to get a valid measure of the effect of source.

Through several metrics of usage and perception, we statistically test each of our five \emph{preregistered} hypotheses.
For instance, one of the aspects we focus on is \textit{utilization}, which entails user's decision to accept or reject attributes provisioned as guidance. Examination of utilization (i) is a less attended but important part of guidance and (ii) draws relevant concepts from the extant judgment and decision making (JDM) literature~\cite{mellers1998judgment}--across psychology and economics--into guidance.
Our emphasis on utilization differs from questions of trust in guidance, especially when coming from AI, which have received considerable attention in research~\cite{liu2022will, fehr2005neuroeconomic, bansal2019case, elish2019moral, green2022flaws, koulu2020human}. We recognize that trust plays an important role. However, the research investigating trust depends largely on stated preference (\textit{what is said}) by users, which may not match with their revealed preference (\textit{what is done})~\cite{engstrom2018demand}. Instead, utilization is about what users do and gives a more objective measure of effect, while also complementing findings about trust--hence our focus on it. 

Another aspect, \textit{regret}, is germane to JDM~\cite{weber2009mindful,loomes1982regret,bell1982regret,zeelenberg1998experience} although largely ignored in guidance. \emph{``[R]egret arises from comparing an obtained outcome with a better outcome that might have occurred had a different choice been made; that is, regret stems from bad decisions''}~\cite[pp.~222]{zeelenberg1998experience}. Moreover, regret is also a function of users' expectations, formed from multiple reference points~\cite{lin2006multiple}. The importance of regret for guidance systems is threefold: (i)~any post-task regret may impact continued utilization of guidance; (ii)~anticipatory regret may distort use of guidance~\cite{tzini2018role}; and (iii)~managing expectations of users may be necessary for success. As part of our study, we measure post-task regret and examine its relation to other guidance characteristics and to users' expectations from guidance. 

Our study findings reveal that: (i) guidance benefits users \textit{during} analysis, although differentially across sources, supporting our thesis; (ii) utilization of guidance varies across stages of analysis--early on, throughout, or later during analysis; (iii) users verify guidance differently across sources; (iv) pre-task \textit{perception} of guidance is adversely affected when attributed to source, although the \textit{decrease} in scores post-task, favors attributed-guidance, pointing to subtleties of source-attribution; (v) users' higher post-task regret by relying on guidance from \ai{} complements the findings about the lower ex-ante trust from \ai{}. At the same time, users find higher benefit of increased confidence from \ai{}'s guidance, calling for a nuanced interpretation of \ai{}'s role. These findings strongly indicate the need to further understand how different guidance sources impact user behavior for designing effective guidance systems.

Our primary contributions include:
\begin{enumerate}[nosep]
    \item Highlighting the importance of \emph{source-attribution} in guidance (via the newly introduced ``from whom'' dimension).
    \item A controlled study with head to head comparison of \textit{three} widely studied sources of guidance: \ai{}, \expert{}, \group{}.
    \item Conceptualizing guidance from a \emph{utilization} point of view, offering objective metrics of measurement (uncertainty, verification, utility) that can be adopted in research and practice.
    \item Conceptualizing guidance-taking as a \emph{decision making problem} and introducing and analyzing constructs of regret and expectations to show new ways to study guidance.
    \item Offering a nuanced perspective about guidance from \ai{}, as users can show both higher post-task benefit and higher post-task regret.
\end{enumerate}


\section{Related Work}
\label{section:relatedwork}

\subsection{Visual Analytics and Guidance}
Visual Analytics (VA) is a human-in-the-loop approach that combines automated analysis techniques with interactive visualizations for an effective understanding, reasoning, and decision-making based on large, complex datasets~\cite{keim2008visual}.
VA systems incorporate concepts from mixed-initiative systems to \emph{``enable users and intelligent agents to collaborate efficiently''}~\cite{horvitz1999principles} by taking initiatives on behalf of each other during analysis. More recently, VA systems have embraced a `human~\emph{is} the loop' perspective--which emphasizes the central role of the user--by enabling the system to implicitly infer their workflow(s), and seamlessly integrating analytics into it~\cite{endert2012semanticinteraction}.
However, automated actions by the system can be mistimed or a result of misinterpreted user intent; whereas, users may need to provide feedback on these automated actions or configure (feedforward) their intent to the system upfront, requiring a continuous, effective dialogue between the two to ensure smooth and effective analytic progress.
Guidance is one such computer-assisted process that aims to actively resolve this ``knowledge gap'' between the user's understanding and the system's capabilities during an interactive analysis session~\cite{ceneda2016characterizing, ceneda2017amending, collins2018guidance}.
In addition, guidance also aims to ensure effective system operation~\cite{smith1986guidelines}, enhance usability~\cite{dix2003human}, improve analysis efficiency, validate insights, build confidence, prevent bias, and improve clarity of findings~\cite{collins2018guidance}.

There have been several approaches to conceptualize and apply guidance in visual analytics to improve the quality of interactions between users and systems.
Engels~\cite{engels1996planning} characterized guidance into a ``what'' dimension (that defines the problem) and a ``how'' dimension (that defines mechanisms to solve the problem). 
Pérez-Messina~et~al.~\cite{perez2022typology} proposed a typology of guidance tasks covering the ``why,'' ``how,'' ``what,'' and ``when'' aspects of guided interactions.
Ceneda~et~al.~\cite{ceneda2016characterizing, ceneda2017amending} characterized guidance based on the user's knowledge gap, the input and output of the guidance process, and its degree, later formalizing a methodology for designing effective guidance systems~\cite{ceneda2020guide}. 
Sperrle~et~al.~\cite{sperrle2020learning, sperrle2021co, sperrle2021topicmodelrefinement} introduced the concept of co-adaptive guidance wherein the user and the system teach and learn from one another during visual data analysis, later contributing a practical framework for developers to build custom guidance strategies~\cite{sperrle2022lotse}.
In this work, we explore an underexplored dimension of guidance, ``from whom'', focusing on the source of guidance, specifically an \ai{}, a human \expert{}, or a \group{} of analysts, and how it impacts a user's perception and usage of guidance.

\subsection{Studies on Utilization of and Reliance on Guidance}
Understanding how guidance is utilized or relied on is critical for designing effective guidance systems, and has also been a key focus of many prior studies.
For instance, 
Wall~et~al.~\cite{wall2021lrg}, Narechania~et~al.~\cite{narechania2021lumos}, and Paden~et~al.~\cite{paden2024biasbuzz} studied how presenting visual traces of a user's interaction history during analysis can help increase their awareness of analytic behavior and mitigate exploration biases.
Sperrle~et~al.~\cite{sperrle2024wizardofozguidance} conducted a Wizard of Oz study to investigate the interaction dynamics between users and systems in co-adaptive guidance scenarios, focusing on the impact of guidance timing, contextualization, and adaptation, as well as the effects of misguidance on user confidence.
Sperrle~et~al.~\cite{sperrle2021learning} also studied how context-dependent user preferences and feedback during topic model refinement can enable the system (not the user) to learn and adapt its subsequent guidance, fostering effective co-adaptive guidance and human-machine collaboration.

In terms of how people utilize and rely on guidance recommendations originating from different sources, we found a complex interplay between human and algorithmic judgments.
Some studies found that users trust and rely on human partners more than AI~\cite{liu2022will, fehr2005neuroeconomic, bansal2019case, elish2019moral, green2022flaws, koulu2020human}, whereas some others found the opposite~\cite{logg2019algorithm, lu2021human, gajos2022people}.
For example, Logg~et~al.~\cite{logg2019algorithm} found people often trust algorithms more than human expertise, despite not fully understanding the algorithm's intricacies.
Several studies found that people's reliance on AI depends on various contextual factors, such as their AI literacy~\cite{jacobs2021machine}, domain expertise~\cite{gaube2021ai}, and amount of feedback~\cite{gajos2022people}.
For example, Gajos~et~al.~\cite{gajos2022people} found that people make more accurate decisions by actively engaging with detailed explanations of AI recommendations--rather than just viewing the recommendations.
Among human partners, studies have revealed differences between guidance from experts versus groups.
Chen~et~al.~\cite{chen2008herd} show that online book purchasing decisions are heavily influenced by consumer recommendations rather than expert opinions. Similarly, Vedadi~et~al.~\cite{vedadi2021herd} find that in information security decisions, users tend to imitate others when faced with uncertainty, impacting their choices more than their personal assessments. This limitation effect extends to software adoption, where user reviews influence lower-ranked products' adoption but not top ones~\cite{duan2009informational}.

In our study, we examine the utilization and perception of guidance from an \ai{}, a human \expert{}, or a \group{} of analysts as source, which represents what users \textit{do}, complementing prior work that investigates trust and reliance, which represent what users \textit{say}. Our study also collects \remove{verbal}\add{typed textual} responses about trust and reliance to make the set of measures comprehensive in having both objective measures of utilization and subjective metrics of perception. We next describe relevant metrics available in prior work that quantify users' utilization of and reliance on guidance.

\subsection{Metrics to Model Utilization of and Reliance on Guidance}
Several metrics have been proposed that measure people's reliance on AI guidance, quantifying people's ``agreement'' and ``disagreement'' with AI recommendations~\cite{lu2021human}, people's ``acceptance'' of incorrect AI recommendations~\cite{buccinca2021trust}, 
people's ``change'' in behavior based on AI recommendations~\cite{kim2023algorithms, logg2019algorithm}, 
and people's propensity to ``delegate'' eventual decision-making to AI~\cite{chiang2021you}. 
For instance, 
Lu~et~al.~\cite{lu2021human} proposed the ``agreement'' and ``disagreement'' metrics, which assess how often user predictions are the same as, or different from, AI recommendations, respectively, when users make predictions before seeing AI recommendations.
Buccinca~et~al.~\cite{buccinca2021trust} measured how often users accept incorrect AI recommendations and how often users make mistakes when their predictions differ from the AI's, with both the user's prediction and the AI's recommendation being wrong.
Kim~et~al.~\cite{kim2023algorithms} proposed ``Switch Fraction'' to assess how often users completely change their answers to match AI recommendations~\cite{kim2023algorithms}.
Logg~et~al.~\cite{logg2019algorithm} proposed ``Weight of Advice'' (WOA) to measure the proportional change in user predictions relative to the change in AI recommendations. 
Chiang~et~al.~\cite{chiang2021you} proposed the ``delegation'' metric to measure how often users let an AI system fully make decisions on their behalf.
These metrics provide a framework for understanding user interactions with guidance systems. In our study, we task participants to select relevant attributes from unfamiliar datasets, but there is no known ground truth in terms of number of attributes to select. Thus, we created new metrics based on these existing ones that better fit our study design.

\subsection{Data Preparation and Subset Selection in Visual Analytics}
Data preparation involves analyzing the data to ensure high-quality results through collection, integration, transformation, cleaning, reduction, and discretization~\cite{dataprep2003zhang}.
As organizations follow a ``load-first'' philosophy and ``dump'' their data into centralized repositories~\cite{gitelman2013raw}, the volume of data often overwhelms users, creating challenges in data navigation, discovery, and monitoring~\cite{deng2017data,fernandez2018aurum, nargesian2020organizing, nargesian2019datalakemanagement}. 
To mitigate these challenges, prior work has utilized several techniques based on the raw data~\cite{deng2017data, miller2018making}, meta-data~\cite{halevy2016goods, monosi}, and users' queries~\cite{brackenbury2018draining,cafarella2009data,zhang2020finding}.
For example, Goods infers metadata from billions of datasets within an organization, making them searchable using keywords~\cite{halevy2016goods}.
Similarly, there exist several proprietary~\cite{montecarlodata, bigeye, datafold} and open-source~\cite{sqllineage,datafold} tools that provide data profile, quality, and lineage information for data observability, monitoring, and pipeline optimization.
For example, 
Profiler~\cite{kandel2012profiler} uses data mining to automatically detect quality issues in tabular data and offers coordinated visualizations for context; whereas, Tableau Prep~\cite{tableauprep}, OpenRefine~\cite{ham2013openrefine}, and Wrangler~\cite{kandel2011wrangler} provide interactive affordances to explore, clean, structure, and shape the data before analysis.

Our study focuses on subset selection\add{~\cite{li2017feature}}, which can be achieved through ``feature set reduction'' to decrease the number of attributes or ``sample set reduction'' to reduce the number of records. Feature set reduction is often used in machine learning to eliminate irrelevant features~\cite{featureselection2015jovic} or perform dimensionality reduction~\cite{fodor2002survey}; whereas sample set reduction is commonly applied in market segmentation~\cite{tynan1987market} to identify specific consumer groups.
Several existing tools facilitate this process of subset selection.
For example, DataPilot~\cite{narechania2023datapilot} presents quality and usage information to assist users in selecting effective subsets from large, unfamiliar tabular datasets; 
DataCockpit~\cite{narechania2023datacockpit} extends these capabilities to multiple relational databases via an open-source Python toolkit.
SmartStripes~\cite{may2011guiding} uses automated filter algorithms and statistical correlation measures along with interactive visualizations.
Notably, visualizing how a selected subset compares to the original dataset has been shown to mitigate selection biases~\cite{gotz2016adaptive, borland2019selection}.
The task in our study is similar to DataPilot's~\cite{narechania2023datapilot}, described next.

\section{Task: Data Preparation and Subset Selection for Visual Analytics}

In this section, we describe the source conditions of our study, the study prototype, the study task, and how we ensure internal validity.

\subsection{Source (Study) Conditions}
We designed a between-subjects study wherein participants were randomly assigned to one of five experimental conditions. In each condition, guidance was presented as if it came from an attributed-source or from an unattributed-source, holding the quality of guidance same across all guidance sources. The five experimental conditions along with their definitions are as follows:

\begin{itemize}[left=46pt]
    \aptLtoX{\item []{\textcolor{colorai}{\textbf{AI}}}}{\item [\ai{}]} Guidance comes from an AI model trained on large information for data analysis tasks.
    \aptLtoX{\item []{\textcolor{colorexpert}{\textbf{Expert}}}}{\item [\expert{}]} Guidance comes from a human expert analyst who is well regarded in industry for acumen in data analysis.
    \aptLtoX{\item[]{\textcolor{colorgroup}{\textbf{Group}}}}{\item [\group{}]} Guidance comes from a group of human analysts in your organization well versed in data analysis.
    \aptLtoX{\item []{\textcolor{colorguidance}{\textbf{Unattributed}}}}{\item[\guidance{}]} Guidance comes without an explicit mention of a source.
    \aptLtoX{\item []{\textcolor{colorcontrol}{\textbf{Control}}}}{\item [\control{}]} No guidance is available in the interface.
\end{itemize}

\subsection{Study Prototype}
\begin{figure*}[ht!]
    \centering
    \includegraphics[width=\linewidth]{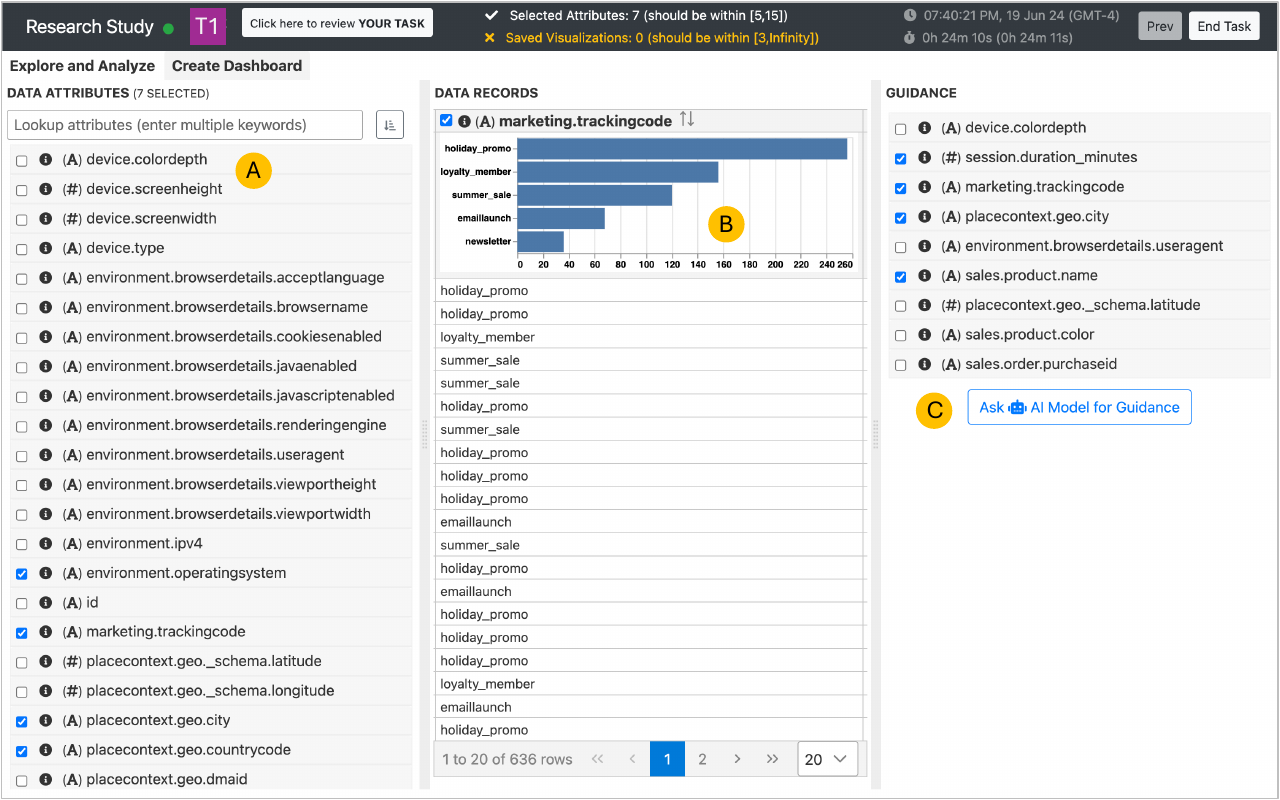}
    \caption{Tab 1 of the study prototype to help users analyze a dataset and select attributes for subsequent use in Tab 2.}
    \label{fig:ui-tab-1}
    \Description{Tab 1 of the study prototype to help users analyze a dataset and select attributes for subsequent use in Tab 2.}
\end{figure*}

\begin{figure*}[ht!]
    \centering
    \includegraphics[width=\linewidth]{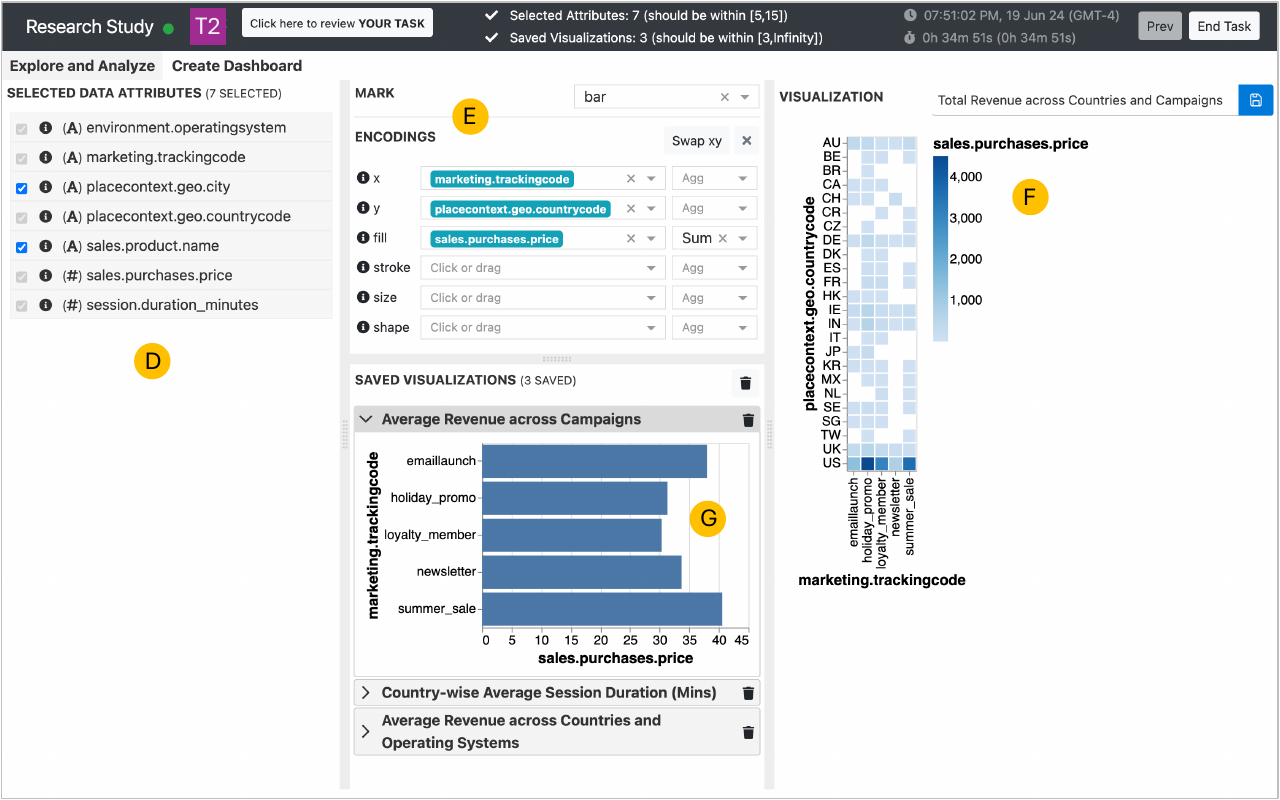}
    \caption{Tab 2 of the study prototype to help users create and save visualizations using the attributes selected from Tab 1.}
    \label{fig:ui-tab-2}
    \Description{Tab 2 of the study prototype to help users create and save visualizations using the attributes selected from Tab 1.}
\end{figure*}

With no existing system that satisfies our goals of experimental control and quality control of guidance, we developed one wherein users can inspect a dataset, select relevant attributes, request guidance, and create visualizations. 
We used Angular~\cite{angular} as our frontend framework and interfaced it with a Python~\cite{python} backend over the REST~\cite{richards2006representational} and websocket~\cite{fette2011websocket} protocols. 
We persisted logged user interactions on the cloud via Google Cloud Logging~\cite{googlecloudlogging}.
Figure~\ref{fig:ui-tab-1} and Figure~\ref{fig:ui-tab-2} show the UI and Table~\ref{table:userinteractions} lists the interactions that are tracked in the UI, described next.

\paragraphHeadingSpace\bpstart{UI Tab 1: Explore and Analyze} 

\begin{itemize}
    \item[(A)] \textbf{Data Attributes} shows the list of dataset attributes along with their definitions and (de)select affordances.
    \item[(B)] \textbf{Data Records} shows the entire dataset in an interactive table along with sorting and pagination affordances.
    \item[(C)] \textbf{Guidance} shows one attribute (up to 10) each time the user requests guidance via a button.
\end{itemize}

\noindent In this tab, users can \textbf{Search} attributes by keywords or \textbf{Sort} attributes (by their name) to organize their search space. For each visible data attribute, users can \textbf{Inspect} it to see a brief description, \textbf{Click} on it to see detailed records, \textbf{Select} it to be in their subset, \textbf{Deselect} it from their subset, and \textbf{Request Guidance} (except for \control{} participants who do not receive any guidance). For the data records, users can \textbf{Paginate} or \textbf{Sort} the interactive table.

\paragraphHeadingSpace\bpstart{UI Tab 2: Create Dashboard}

\begin{itemize}
    \item[(D)] \textbf{Selected Data Attributes}, similar to (A)~\textbf{Data Attributes}, shows the list of attributes \emph{selected} by the user.
    \item[(E)] \textbf{Mark and Encodings} shows affordances to create visualizations: a visualization mark type (bar, point, line), visual encodings (x, y, fill color, size, shape), and encoding aggregations (sum, mean, max, min).
    \item[(F)] \textbf{Visualization} shows the visualization based on the \textbf{Mark and Encodings} along with an affordance to save it.
    \item[(G)] \textbf{Saved Visualizations} shows the list of all saved \textbf{Visualization}s, including affordances to delete one or all.
\end{itemize}

\noindent In this tab, users can \textbf{Inspect} and \textbf{Deselect} selected attributes from the Selected Data Attributes view. Additionally, users can \textbf{Create} visualizations by changing the mark type, changing and resetting visual encodings and aggregations, and swapping the attributes mapped to xy axes.
Lastly, users can \textbf{Save} visualizations or \textbf{Delete} one or all saved visualizations.

\begin{table*}[!ht]
    \centering
    \begin{tabular}{rlllc}
    
    & \textbf{Interaction Name} & \textbf{Description} & \textbf{View(s)} & \textbf{Cue} \\
    \midrule

    1 & Search attributes & Filters the list of attributes based on user-input search term. & A & \faFilter \\

    2 & Sort attributes & Toggle sorts the attributes by their name. & A & \faSortAmountAsc \\
    
    \begin{imageonly}\circlednumber{3}\end{imageonly} & Inspect attribute & Shows a brief description about the attribute in a tooltip. & ABCD & \faInfoCircle \\
    
    \begin{imageonly}\circlednumber{4}\end{imageonly} & Click attribute & Shows an overview of the attribute's data distribution and detailed records. & AC &  \\

    \begin{imageonly}\circlednumber{5}\end{imageonly} & Select attribute & Selects an attribute to be in the desired subset. & ABC & \faCheckSquare \\
    
    \begin{imageonly}\circlednumber{6}\end{imageonly} & Deselect attribute & Removes a previously selected attribute from the subset. & ABCD & \faSquareO \\
    
    \begin{imageonly}\circlednumber{7}\end{imageonly} & Paginate datatable & Paginates through different pages of the datatable. & B & \faAngleDoubleLeft~\faAngleDoubleRight \\
    
    \begin{imageonly}\circlednumber{8}\end{imageonly} & Sort records & Toggle sorts the datatable records by that attribute. & B & \faSort \\
    
    \begin{imageonly}\circlednumber{9}\end{imageonly} & Request guidance & Requests an attribute recommendation as guidance (a maximum 10 times). & C & \\
        
    10 & Change tab & Switch between the ``Explore and Analyze'' and Create Dashboard'' tabs. & T1--T2 & \\
    
    11 & Change mark type & Changes mark type (e.g. bar, point) of the visualization. & E & \\
    
    \begin{imageonly}\circlednumber{12}\end{imageonly} & Change encoding & Assigns (or clears) an attribute to a visual encoding. & E & \\
    
    \begin{imageonly}\circlednumber{13}\end{imageonly} & Change aggregation & Assigns (or clears) the encoding's aggregation function (e.g. max). & E & \\
    
    \begin{imageonly}\circlednumber{14}\end{imageonly} & Swap xy axes & Swaps the attributes encoded to the x-axis and y-axis. & E & \\
    
    15 & Reset encodings & Revert to the default encoding for visualizations. & E & \faTimes \\
    
    \begin{imageonly}\circlednumber{16}\end{imageonly} & Save vis & Saves the current visualization. & F & \faSave \\
    
    \begin{imageonly}\circlednumber{17}\end{imageonly} & Delete one vis & Remove a specific visualization from the saved list. & G & \faTrash \\
    
    18 & Delete all vis & Remove all visualizations from the saved list. & G & \faTrash \\
    \bottomrule
    
    \end{tabular}
    \caption{User interactions tracked in the study prototype including their name (\textbf{Interaction Name}), a short \textbf{Description} and the \textbf{View(s)} (and icon \textbf{Cues}) they occur in (Figures~\ref{fig:ui-tab-1},~\ref{fig:ui-tab-2}). The interactions with \textcolor{orange}{orange serial numbers} (e.g, \begin{imageonly}\circlednumber{5}\end{imageonly} Select attribute) can be directly mapped to one or more attributes at any time; an aspect that will be utilized in analysis (Section~\ref{section:results}).}
    \label{table:userinteractions}
\end{table*}


\subsection{Main (Study) Task}
\label{sec:maintask}

Using the study prototype, we tasked participants to perform the following analysis task:

\begin{quote}
    \emph{Analyze the provided dataset of attributes derived from online customer behaviors. Design multiple visualizations indicating meaningful drivers of dollar (\$) sales revenue for the company.}

    \vspace{0.5em}
    
    \emph{Select between 5 to 15 attributes (both inclusive); and make at 3 visualizations. It is important for you to pay attention to the attributes you select and the visualizations you create from them.}

    \vspace{0.5em}

    \emph{Note that these attributes and visualizations will be examined by your boss, who will then decide whether to use them for their own report. Spend around 20 minutes for this task.}
\end{quote}

For the conditions that received guidance (\ai{}, \expert{}, \group{}, \guidance{}) there were additional instructions:

\begin{quote}
    \emph{To help you perform this complex task, guidance is on the way. This guidance may be helpful to perform the task well. The guidance will appear on the right-hand side panel on the UI. Your judgment is important in accepting or in rejecting any guidance you receive. It is your decision whether and which guidance to use. You will receive guidance:}

    \begin{enumerate}[nosep, left=6em]
        \aptLtoX{\item[]\textcolor{colorai}{\textbf{AI}}}{\item [(For \ai{})]} from an AI model trained on large information for data analysis tasks.
        \aptLtoX{\item[](For \textcolor{colorexpert}{\textbf{Expert}})}{\item [(For \expert{})]} from a highly regarded industry expert in data analysis.
        \aptLtoX{\item[](For \textcolor{colorgroup}{\textbf{Group}})}{\item [(For \group{})]} from a group of data analysts in your organization well versed in data analysis.
        \aptLtoX{\item[](For \textcolor{colorguidance}{\textbf{Unattributed}})}{\item[(For \guidance{})]} that may be helpful to perform the task well.
    \end{enumerate}
        
\end{quote}

We note that our primary focus is on participants' selection of relevant data attributes, aligned with our study goals and hypotheses. Our analyses, presented later, emanate from this primary focus. To motivate participants' decision making toward selection of attributes, their task also involves creating and saving visualizations using their selected attributes. The visualization outcomes are not meant to be evaluated since those have inherently subjective, ``in-the-eyes-of-the-beholder'' nature, making it hard to meaningfully evaluate as task performance. As well, those do not directly address our hypotheses.

\subsection{Internal Validity}
\label{sec:internalvalidity}

Concerned with the internal validity of the study, we hold the quality of guidance constant across the four guidance conditions: \guidance{}, \ai{}, \expert{}, and \group{}. 
Of the total 33 attributes in the dataset that participants can select, we carefully identify 14 attributes that are important and relevant to our study task.
From these 14 relevant attributes, we select 7, and then from the remaining 19 irrelevant attributes, we select 3, to create a list of 10 total attributes.
This 7:3 ratio of relevant to irrelevant attributes was intentionally chosen to encourage participants to exercise judgment when using the provided guided attributes, rather than using them indiscriminately (\textit{carte blanche}). 
Our study description emphasizes the importance of this judgment in deciding among guided attributes.
Finally, we randomize the order in which these 10 attributes are recommended in the user interface, to enforce experimental validity by avoiding an order effect. 
By using the same proportion of relevant and irrelevant attributes that are guided, we achieve internal validity. However, to the extent the quality of guidance can vary with source, we lose some external validity. To achieve both internal and external validity requires a much larger study, which is beyond the scope of this paper.


\section{Hypotheses}
\label{section:hypotheses}


\label{sec:hypotheses}

A primary goal of guidance is to minimize the ``knowledge-gap'' between the user and the system to enhance analytic workflows~\cite{collins2018guidance,ceneda2016characterizing}. 
Received wisdom suggests that the framework of guidance for visual analytics is commonly conceptualized agnostic of the source of guidance~\cite{ceneda2019review,ceneda2020guide}. In guidance without source-attribution, the focus for the beneficiary-user is on the guidance provided in the system, devoid of the entity behind the guidance. Thus, we find it useful to establish user behavior with this kind of source-agnostic guidance, termed, \guidance{} guidance. Comparing users' behaviors in \guidance{} guidance with that of the condition of no-guidance, or \control{}, allows a direct assessment of the impact of guidance on users' behaviors, un-confounded by any source-attribution (i.e., \ai{}, \expert{}, \group{}). This comparison is the basis of the first three hypotheses. Only then, we examine the incremental impact of source-attribution in a set of two additional hypotheses. Below, we discuss the source-agnostic hypotheses followed by the source-specific hypotheses.
All five hypotheses were preregistered and can be accessed at \remove{\url{https://osf.io/q6ca5}}\add{\url{https://osf.io/q6ca5}}.

\subsection{Source-Agnostic Hypotheses about Guidance}
The three hypotheses in this section compare \guidance{} guidance to no guidance (\control{}). 
The premise of guidance is that a user has a knowledge gap with respect to the system being used. Knowledge gap manifests in higher uncertainty the user faces when considering whether to select certain attributes. Our first hypothesis, \textbf{H1}, proposes that \guidance{} guidance can aid in uncertainty reduction relative to the \control{} condition. The reduction in uncertainty under guidance gives higher confidence in selection of attributes, which can result in selecting more number of attributes for the task at hand, especially when those attributes come from guidance. Thus, our second hypothesis, \textbf{H2}, proposes that more attributes are likely to be selected under \guidance{} guidance as compared to \control{}. Also, the reduction in uncertainty makes decision making about attribute selection quicker, giving our third hypothesis, \textbf{H3}, that total time for task-completion is likely to be lower in \guidance{} guidance relative to \control{} condition. We next present objective metrics that form the basis for statistically testing these hypotheses.

\aptLtoX{\begin{itemize}[left=0pt]
    \item [\textbf{H1}] \textbf{Participants who receive guidance will \emph{be less uncertain about their attribute selections}.}
    
    \bpstart{Metric(s).} We define \emph{uncertainty} as the variance in the number of interactions with attributes. A higher variance indicates greater uncertainty, as it suggests the participant is repeatedly interacting with an attribute, likely due to difficulty in deciding whether to select it or not. 

    \begin{equation}
        \text{Uncertainty} = \frac{1}{n} \sum_{i=1}^{n} (x_i - \mu)^2
        \label{eq:hypothesis1metric1}
    \end{equation}

    where $n$ $=$ total number of attributes in the dataset, $x_i$ $=$ number of interactions with attribute $i$, and $\mu = \frac{1}{n} \sum_{i=1}^{n} x_i$ = mean number of interactions across all attributes.

    \item [\textbf{H2}] \textbf{Participants who receive guidance will \emph{select more attributes}.}
    
    \bpstart{Metric(s).} Attribute selection refers to shortlisting an attribute to include in the final phase of task.

    \item [\textbf{H3}] \textbf{Participants who receive guidance will \emph{take lesser time} to complete the task.}
    
    \bpstart{Metric(s).} We measure \emph{time} in two ways: 
    
    \begin{itemize}[nosep]
        \item[(a)] \emph{Duration of the task (in minutes).}
        \item[(b)] \emph{Total number of interactions performed during the task.}
    \end{itemize}

    The duration measures the total time spent performing the task, while the number of interactions is proxy for the amount of engagement with the UI.

\end{itemize}}{\begin{enumerate}[left=0pt]
    \item [\textbf{H1}] \textbf{Participants who receive guidance will \emph{be less uncertain about their attribute selections}.}
    
    \bpstart{Metric(s).} We define \emph{uncertainty} as the variance in the number of interactions with attributes. A higher variance indicates greater uncertainty, as it suggests the participant is repeatedly interacting with an attribute, likely due to difficulty in deciding whether to select it or not. 

    \begin{equation}
        \text{Uncertainty} = \frac{1}{n} \sum_{i=1}^{n} (x_i - \mu)^2
        \label{eq:hypothesis1metric1}
    \end{equation}

    where $n$ $=$ total number of attributes in the dataset, $x_i$ $=$ number of interactions with attribute $i$, and $\mu = \frac{1}{n} \sum_{i=1}^{n} x_i$ = mean number of interactions across all attributes.

    \item [\textbf{H2}] \textbf{Participants who receive guidance will \emph{select more attributes}.}
    
    \bpstart{Metric(s).} Attribute selection refers to shortlisting an attribute to include in the final phase of task.

    \item [\textbf{H3}] \textbf{Participants who receive guidance will \emph{take lesser time} to complete the task.}
    
    \bpstart{Metric(s).} We measure \emph{time} in two ways: 
    
    \begin{enumerate}[nosep]
        \item \emph{Duration of the task (in minutes).}
        \item \emph{Total number of interactions performed during the task.}
    \end{enumerate}

    The duration measures the total time spent performing the task, while the number of interactions is proxy for the amount of engagement with the UI.

\end{enumerate}}

\subsection{Source-Specific Hypotheses}
Among guidance sources, human \expert{} guidance is valued for its credibility and specialized knowledge~\cite{feltovich2006studies}, while an \ai{} model offers efficient and objective suggestions~\cite{brynjolfsson2016rapid}, and a human \group{} provides diverse perspectives and collective expertise~\cite{page2008difference}. These findings lead us to propose \textbf{H4} about relative utilities from three sources. However, each source has limitations: human \expert{} guidance may be biased or be involved in tension with novices~\cite{thom2010you}, \ai{} model may lack contextual understanding and interpretability~\cite{ribeiro2016should}, and human \group{} guidance can suffer from group-think and inconsistency~\cite{banerjee1992simple}. Synthesizing these results, we call out verification of provided guidance and propose \textbf{H5}. Previous studies have shown varied perceptions of guidance based on its source and context, with mixed reports on the usefulness of human versus AI guidance~\cite{liu2022will,fehr2005neuroeconomic,logg2019algorithm} and expert versus group guidance~\cite{banerjee1992simple,chen2008herd}. Moreover, these studies were performed under different platforms ranging from emails to scenario-based visual analytics, and outcomes were measured as stated responses on semantic scales. No single study we know of compares all three sources - \ai{}, \expert{}, \group{}, nor presents comparison on a common platform, nor measures utilization, nor uses objective metrics of actions and interactions. 
Accordingly, we present objective metrics that form the basis for statistically testing these hypotheses, while also supplementing them with subjective responses on semantic scales.

\begin{itemize}[left=0pt]

    \aptLtoX{\item []\textbf{H4}}{\item [\textbf{H4}]} \textbf{Participants will find guidance to \emph{have more utility} when it comes from \expert{} > \ai{} > \group{} ($>$ implies more).}

    \bpstart{Metric(s).} We measure \emph{utility} as the total number of attributes selected at the end of the task.

    Drawing from utility theory for decision making~\cite{fishburn1979utility}, the construct \textit{utility} assesses a user's preference for outcome. We seek a quantitative representation of a qualitative preference for attributes, which in turn is pegged to the task. To accomplish the task in our study, a user explores and reviews attributes, both on their own and with help from guided attributes, before finalizing the attributes. We thus use the total number of attributes selected as the metric of utility to test \textbf{H4}. Note that: (i) users did not know about our hypotheses and had no incentive to select few or more attributes than their judgment; (ii) the task's goal was for users to select only those attributes that they judged to be relevant; (iii) results show that users did not select too few or too many attributes. That said, as the literature of over 100 years testifies, measurement of utility, a subjective construct, does not lend itself to a unique measure~\cite{moscati2018measuring}. One can design more involved studies solely to measure utility, but is outside the scope of this paper~\cite{moscati2018measuring}.   

    \aptLtoX{\item []\textbf{H5}}{\item [\textbf{H5}]} \textbf{Participants will \emph{verify the guidance more} when it comes from \ai{} > \expert{} > \group{} ($>$ implies more).}
    
    \bpstart{Metric(s).} We define \emph{verification} as the change in users' attention (a) towards guided attributes before and after receiving them as guidance and (b) between guided and unguided attributes, providing a comprehensive assessment of how guidance sources influence verification behavior.

    \begin{itemize}[nosep]
        
        \item[(a)] \emph{Difference in number of interactions with guided attributes after and before receiving them as guidance,} or \(\Delta x\): 

        \begin{equation}
            \Delta x = \sum_{i=1}^{k} (x_{i, \text{after}} - x_{i, \text{before}})
            \label{eq:hypothesis5metric1}
        \end{equation}
        
        where: \( x_{i, \text{after}} \) and \( x_{i, \text{before}} \) are the number of interactions with guided attribute \( i \) after and before receiving as guidance, respectively, and \( k \) is the total number of guided attributes.

        Since there may be zero interactions with an attribute before and after receiving it as guidance, we model this metric as a \emph{difference} instead of \emph{ratio}, avoiding divide-by-zero errors and enabling statistical comparisons.

        \item[(b)] \emph{Ratio of number of interactions with guided attributes to total number of interactions with all attributes,} or \(R\):
        
        \begin{equation}
            R = \frac{ \sum_{i=1}^k x_{i, \text{guided}} }{ \sum_{j=1}^n x_{j, \text{total}} }
            \label{eq:hypothesis5metric2}
        \end{equation}

        where: \( x_{i, \text{guided}} \) is the number of interactions with guided attribute \( i \), \( k \) is the total number of guided attributes, \( x_{j, \text{total}} \) is the number of interactions with all attributes, and \( n \) is the total number of attributes.
        Since users may not interact with all attributes, particularly in datasets with many attributes (e.g., thousands or even 33 as in our case), and considering that our guidance was capped at 10 attribute suggestions, we model this metric as a \emph{ratio} instead of \emph{difference}. Unlike (a), this ratio avoids divide-by-zero errors and provides a clearer comparison of the user's attention on guided attributes relative to all attributes.
    \end{itemize}
\end{itemize}

\section{User Study}
\label{section:evaluation}

\subsection{Participants}
We recruited participants by emailing relevant mailing lists within a public U.S. university.
Each interested applicant was screened based on their self-reported visualization literacy (at least 3 out of 5), age (at least 18 years), and physical location (in the U.S.), as per our study protocol approved by the ethics board.
We invited 109 screened applicants to participate in the study of whom 90 participants completed the study. We discarded data of three of these participants from the resultant analysis due to suspicious/insufficient interaction behavior, resulting in 87 valid participants\footnote{Participants whose number of interactions or task duration fell outside the mean $\pm$ 2 standard deviations were deemed outliers, and excluded from analysis.}.

Our valid participants were either pursuing or had received \emph{professional}~(n=2), \emph{associates}~(1), \emph{bachelors}~(24), \emph{masters}~(38), \emph{doctoral}~(21) degrees or a \emph{technical certificate} in \emph{computer science}~(n=49), \emph{human-centered computing}~(4), \emph{human-computer interaction}~(8), \emph{mechanical engineering}~(4), \emph{analytics}, \emph{applied and computational mathematics}, \emph{business administration}~(2), \emph{chartered finance analyst}, \emph{cybersecurity}, \emph{digital media}, \emph{economics}, \emph{electrical and computer engineering}, \emph{history}, \emph{industrial engineering}, \emph{information management}, \emph{mechatronics, robotics and automation engineering}, \emph{pharmaceutical sciences}, \emph{physics}, \emph{public health}, \emph{public policy}, \emph{robotics}, and \emph{statistical science}.
Demographically, they were in the \emph{18-24}~(n=47), \emph{25-34}~(38), \emph{35-44}~(1), or \emph{45-54}~(1) age groups in years and of \emph{female}~(42), \emph{male}~(44), or \emph{preferred not to say}~(1) genders. 
They reported their level of experience looking at data in a visual form (e.g., scatterplot, bar chart) on a scale from 1 (non-expert) to 5~(expert): 3~(n=36), 4~(37), 5~(14).
They also reported their level of experience working with analysis tools such as Excel (n=85), Programming~(75), Tableau~(48), PowerBI~(12), QlikView~(3), D3~(3), SQL, SAS, Plotly, Kibana, Colaboratory, .NET, matplotlib, MicroStrategy, Looker, Grafana, Sheets, and Gephi.

\begin{figure*}
    \includegraphics[width=0.98\linewidth]{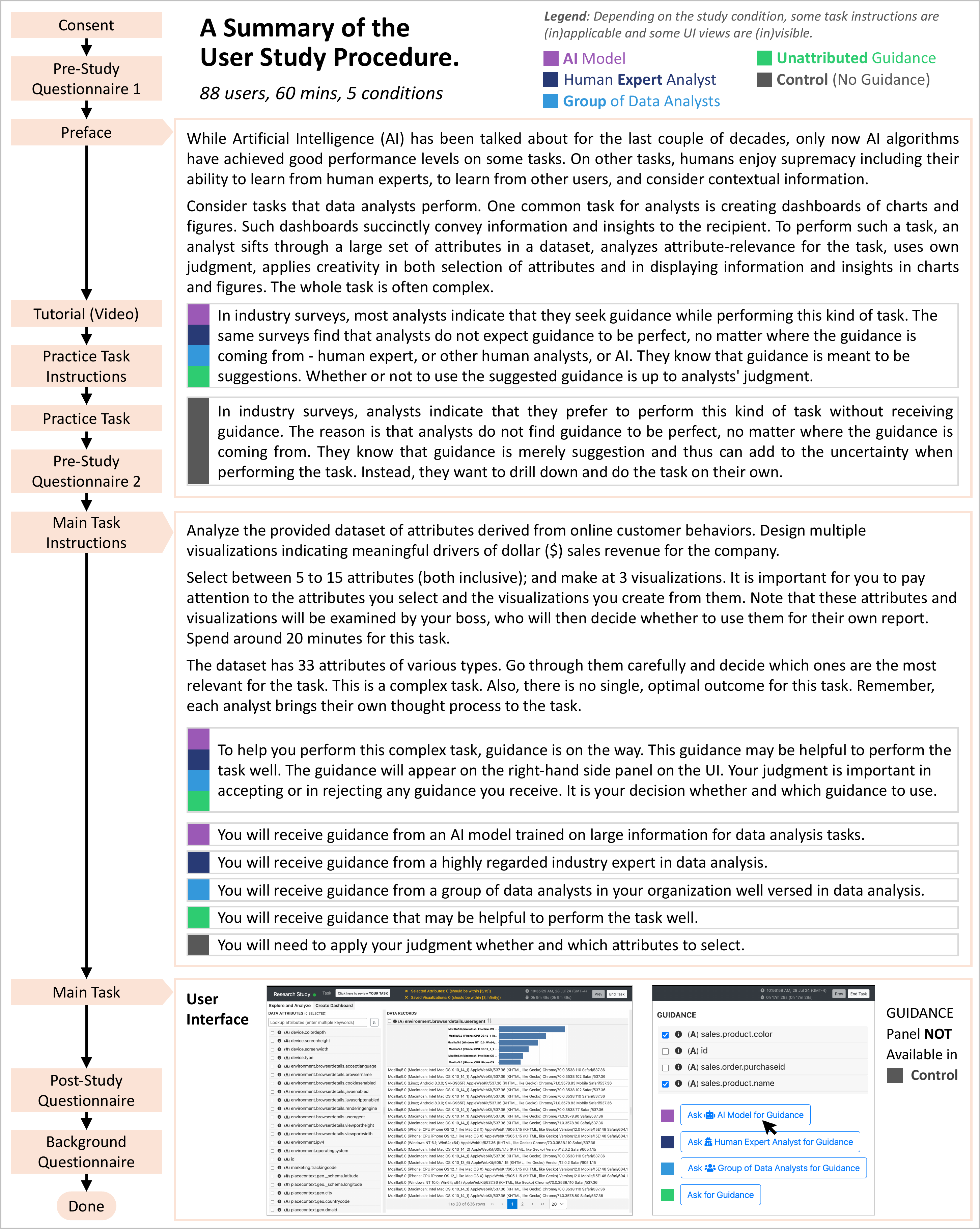}
    \centering
    \caption{Overview of our Study Design including the various steps, task instructions, and screenshots of the study prototype.}
    \label{fig:study-design}
    \Description{Overview of our Study Design including the various steps, task instructions, and screenshots of the study prototype.}
  \end{figure*}

\subsection{Study Session}
We conducted the study remote asynchronously, providing each participant with a unique link to the study's user interface and giving a maximum two weeks to complete the task, albeit in a single, focused session. Each study session lasted between 30 and 60 minutes. We compensated each participant with a \$15 gift card for their time.

During a study session, participants first provided consent (\textbf{Consent}), received relevant background information on providing guidance during data analysis workflows (\textbf{Preface}), and filled a questionnaire to provide their general prior beliefs about and experiences receiving guidance in life (\textbf{Pre-Study Questionnaire 1}).
Next, participants saw a video tutorial (\textbf{Training}) that demonstrated the features of the study interface and performed a practice task (\textbf{Practice Task Instructions} and \textbf{Practice Task}) to familiarize themselves with the interface.
After practice, participants saw the actual task instructions (\textbf{Main Task Instructions}) and filled a questionnaire to provide their prior beliefs with respect to the task (\textbf{Pre-Study Questionnaire 2}).
Next, participants performed the main task using the study interface (\textbf{Main Task}). 
Lastly, participants filled a feedback questionnaire based on the study task (\textbf{Post Study Questionnaire}) and another questionnaire to collect their demographic information (\textbf{Background Questionnaire}). 
Participants' interactions with the interface (e.g., the attributes they selected) were logged during the study for subsequent analysis. Figure~\ref{fig:study-design} illustrates the entire study procedure; whereas detailed study questionnaires are available in supplemental material.


\section{Results}
\label{section:results}


We first describe our choice of statistical analysis, then present findings related to all hypotheses, followed by additional analyses of usage behavior and responses from the pre- and post-study questionnaires. 
We include summary visualizations and statistical findings related to the hypotheses and the questionnaires in the Appendix.

\subsection{Statistical Analysis - Non-parametric}
A common analysis to compare study conditions, as in this paper, is a statistical test for comparison of means of responses, which relies upon the assumption of normal distribution of responses. The number of participants in each of our five conditions is $17$, or $18$, which falls well below the number $30$, at which the distribution of the sample mean of a metric in a condition approaches the normal distribution. Also, the measures in our study are of two types: subjective, stated responses to questions, and objective metrics of usage. The distribution of several metrics, within each condition, depicts considerable skewness, violating assumption of normal distribution. Thus, we use non-parametric analysis and tests, which do not rely on inherent assumptions about distribution and provide robustness to compare conditions for responses and metrics alike. This allows a consistent approach to both sets of measures. In particular, we use \textit{median}, instead of mean, and perform non-parametric statistical tests. Additionally, to comprehensively examine statistical relationships among a large set of responses, we do a parametric, multivariate regression analysis.
We conducted all statistical analyses using well-established Python libraries: scikit-learn\footnote{\url{https://scikit-learn.org/stable/}} and statsmodels\footnote{\url{https://www.statsmodels.org/stable/index.html}}.

\subsection{Results of Source-Agnostic Hypotheses}

We report the results of our three source-agnostic hypotheses (\textbf{H1}--\textbf{H3}) involving \guidance{} and \control{} conditions, using one-sided Mann-Whitney U tests (one-sided as the hypotheses are one-sided). 

\begin{itemize}[left=0pt]
    \aptLtoX{\item []\textbf{H1}}{\item [\textbf{H1}]} \textbf{\guidance{} participants who receive guidance will \emph{be less uncertain about their attribute selections} than \control{} participants who do not receive guidance.} 

    \paragraphHeadingSpace\colorbox{lightergray}{\textbf{Metric:} Variance in the number of interactions with attributes.}        

    \paragraphHeadingSpace\noindent The median variance in the number of interactions across all attributes for \guidance{} participants (median=19.18) was smaller than \control{} participants (median=21.82).
    These results \textbf{\remove{support}\add{directionally favor} our hypothesis}, but are not statistically significant (\emph{p}-value=0.34).

    \aptLtoX{\item []\textbf{H2}}{\item [\textbf{H2}]} \textbf{\guidance{} participants who receive guidance will \emph{select more attributes} than \control{} participants who do not receive guidance.}

    \paragraphHeadingSpace\colorbox{lightergray}{\textbf{Metric:} Total number of attributes selected at task end.}
    
    \paragraphHeadingSpace\noindent\guidance{} participants selected more attributes (median=10.5) than \control{} (median=6.0). These results \textbf{\remove{support}\add{directionally favor} our hypothesis} and are \textbf{statistically significant} (\textbf{\emph{p}-value=0.001}).

    \aptLtoX{\item []\textbf{H3}}{\item [\textbf{H3}]} \textbf{\guidance{} participants who receive guidance will \emph{take lesser time} to complete the task than \control{} participants who do not receive guidance.}

    \paragraphHeadingSpace\colorbox{lightergray}{\textbf{Metric:} Total duration of the task (in minutes).}
        
    \paragraphHeadingSpace\noindent\guidance{} spent less time (median=12.76 minutes) than \control{} (median=13.57 minutes). These results \textbf{\remove{support}\add{directionally favor} our hypothesis} but are not statistically significant (\emph{p}-value=0.67).

    \paragraphHeadingSpace\colorbox{lightergray}{\textbf{Metric:} Total number of interactions made during the task.}
            
    \paragraphHeadingSpace\noindent\guidance{} performed less interactions (median=141) than \control{} (median=145.5). These results \textbf{\remove{support}\add{directionally favor} our hypothesis} but are not statistically significant (\emph{p}-value=0.72).
    
\end{itemize}

\noindent For more details, refer to Table~\ref{table:results-hypotheses-summary-h123} in Appendix~\ref{section:appendix-h1h2h3}.

\subsection{Results of Source-Specific Hypotheses}

We report results of our two source-specific hypotheses (\textbf{H4}--\textbf{H5}) involving \ai{}, \expert{}, and \group{} conditions, using pairwise one-sided Mann-Whitney U tests with Bonferroni correction (one-sided tests as both hypotheses are one-sided).

\aptLtoX{\begin{itemize}
        \item [\textbf{H4}] \textbf{Participants will find guidance to \emph{have more utility} when it comes from \expert{} > \ai{} > \group{}.}

    \paragraphHeadingSpace\colorbox{lightergray}{\textbf{Metric:} Total number of attributes selected at task end.}
    
    \paragraphHeadingSpace\noindent\ai{} (median=9) and \expert{} (median=9) both selected more attributes than \group{} (median=7). 
    These results \textbf{\remove{partially support}\add{directionally favor} our hypothesis\add{, albeit partially}}; the pairwise comparisons between the three conditions revealed \textbf{statistical significance} between \expert{} and \group{} (\textbf{\emph{p}-value=0.01}) and \ai{} and \group{} (\textbf{\emph{p}-value=0.04}).

    Notably, all three source-specific conditions selected fewer attributes than \guidance{} (median=10.5) and more attributes than \control{} (median=6) (\textbf{H2}).

    \item [\textbf{H5}] \textbf{Participants will \emph{verify the guidance more} when it comes from \ai{} > \expert{} > \group{}.}

    \paragraphHeadingSpace\colorbox{lightergray}{\textbf{Metric:} Difference in the number of interactions with guided}
    
    \colorbox{lightergray}{attributes after and before guidance.}

    \paragraphHeadingSpace\noindent\ai{} had the largest difference in number of interactions with guided attributes after and before receiving them as guidance (median=31), followed by \expert{} (median=14), and then \group{} (median=12.5). 
    These results \textbf{\remove{support}\add{directionally favor} our hypothesis} and the pairwise comparisons between the three conditions revealed \textbf{statistical significance} between \ai{} and \group{} (\textbf{\emph{p}-value=0.003}).

    \paragraphHeadingSpace\colorbox{lightergray}{\textbf{Metric:} Ratio of \# interactions with guided attributes to}

    \colorbox{lightergray}{\# interactions with all attributes.}

    \paragraphHeadingSpace\noindent\ai{} had the largest ratio of the number of interactions with guided attributes to number of interactions with all attributes (median=0.36), followed by \expert{} (median=0.23), and then \group{} (median=0.19). 
    These results \textbf{\remove{support}\add{directionally favor} our hypothesis} and the pairwise comparisons between the three conditions revealed \textbf{statistical significance} between \ai{} and \group{} (\textbf{\emph{p}-value=0.01}).

\end{itemize}
}{\begin{itemize}[left=0pt]
    
    \item [\textbf{H4}] \textbf{Participants will find guidance to \emph{have more utility} when it comes from \expert{} > \ai{} > \group{}.}

    \paragraphHeadingSpace\colorbox{lightergray}{\textbf{Metric:} Total number of attributes selected at task end.}
    
    \paragraphHeadingSpace\noindent\ai{} (median=9) and \expert{} (median=9) both selected more attributes than \group{} (median=7). 
    These results \textbf{\remove{partially support}\add{directionally favor} our hypothesis\add{, albeit partially}}; the pairwise comparisons between the three conditions revealed \textbf{statistical significance} between \expert{} and \group{} (\textbf{\emph{p}-value=0.01}) and \ai{} and \group{} (\textbf{\emph{p}-value=0.04}).

    Notably, all three source-specific conditions selected fewer attributes than \guidance{} (median=10.5) and more attributes than \control{} (median=6) (\textbf{H2}).

    \item [\textbf{H5}] \textbf{Participants will \emph{verify the guidance more} when it comes from \ai{} > \expert{} > \group{}.}

    \paragraphHeadingSpace\colorbox{lightergray}{\textbf{Metric:} Difference in the number of interactions with guided}
    
    \colorbox{lightergray}{attributes after and before guidance.}

    \paragraphHeadingSpace\noindent\ai{} had the largest difference in number of interactions with guided attributes after and before receiving them as guidance (median=31), followed by \expert{} (median=14), and then \group{} (median=12.5). 
    These results \textbf{\remove{support}\add{directionally favor} our hypothesis} and the pairwise comparisons between the three conditions revealed \textbf{statistical significance} between \ai{} and \group{} (\textbf{\emph{p}-value=0.003}).

    \paragraphHeadingSpace\colorbox{lightergray}{\textbf{Metric:} Ratio of \# interactions with guided attributes to}

    \colorbox{lightergray}{\# interactions with all attributes.}

    \paragraphHeadingSpace\noindent\ai{} had the largest ratio of the number of interactions with guided attributes to number of interactions with all attributes (median=0.36), followed by \expert{} (median=0.23), and then \group{} (median=0.19). 
    These results \textbf{\remove{support}\add{directionally favor} our hypothesis} and the pairwise comparisons between the three conditions revealed \textbf{statistical significance} between \ai{} and \group{} (\textbf{\emph{p}-value=0.01}).

\end{itemize}}

\noindent For more details, refer to Table~\ref{table:results-hypotheses-summary-h45} in Appendix~\ref{section:appendix-h4h5}.

\subsection{Results of User Behavior Analysis}
Besides metrics for the analysis of hypotheses, we compute other metrics that shine light on users' analytic behaviors while performing the task and can enrich our understanding of utilization of guidance. Specifically, user behavior analyses indicate interesting new hypotheses that draw from the rich psychology literature and set valuable future research directions.

\subsubsection{Exploration of Unique Attributes}

\begin{figure}[H]
    \includegraphics[width=0.8\linewidth]{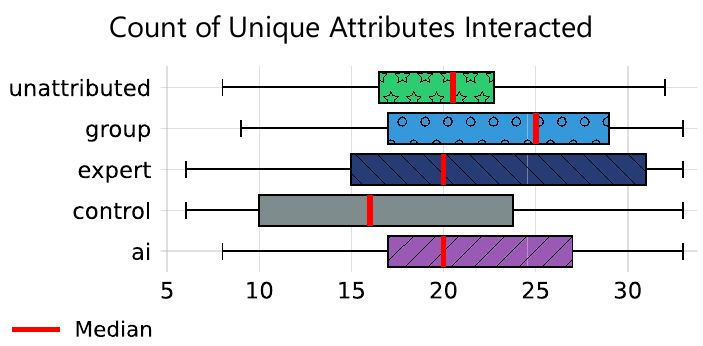}
    \centering
    \caption{Number of unique attributes interacted.}
    \label{fig:log_analysis_count_unique_attributes_interacted}
    \Description{Number of unique attributes interacted.}
\end{figure}

Figure~\ref{fig:log_analysis_count_unique_attributes_interacted} shows the distribution of the total number of unique attributes interacted by users.
\group{} shows the highest exploration~(median=25) followed by \guidance{}~(20.5), \ai{}~(20), \expert{}~(20), and \control~(16). Directionally, among attributed sources, less exploration under \ai{} and \expert{} suggests a belief in more prescriptive guidance from these two sources relative to \group{}.

\subsubsection{Exploitation of Availability of Guidance}

\begin{figure}[H]
    \includegraphics[width=0.8\linewidth]{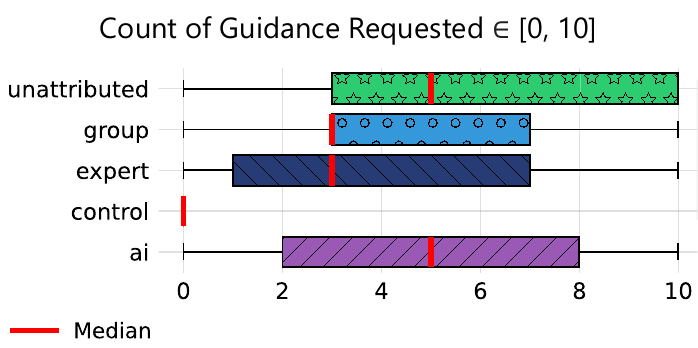}
    \centering
    \caption{Number of times guidance was requested.}
    \label{fig:count-guidance-requested}
    \Description{Number of times guidance was requested across study conditions.}
\end{figure}

Figure~\ref{fig:count-guidance-requested} shows the distribution of the number of times users requested guidance, when available. \guidance{} and \ai{} find the most exploitation~(median=5) followed by \expert{} and \group{}~(median=3).
Four participants in each of the \ai{}, \expert{}, \group{}, and \guidance{} conditions did not request guidance at all. Among attributed sources, these results may indicate a belief of more to be gained from exploitation of \ai{} than from others; that is, more \textit{salience}~\cite{higgins1996activation} of \ai{}.

\subsubsection{Eagerness to Access Guidance}

\begin{figure}[H]
    \includegraphics[width=0.85\linewidth]{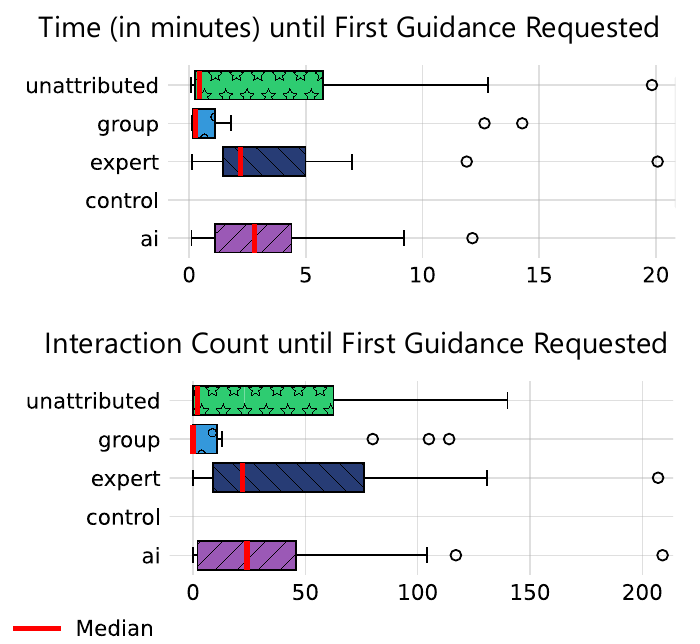}
    \centering
    \caption{Time taken (in minutes) and interaction count until guidance was requested for the first time.}
    \label{fig:log_analysis_interaction_count_duration_until_first_guidance}
    \Description{Time taken (in minutes) and interaction count until guidance was requested for the first time.}
\end{figure}

Figure~\ref{fig:log_analysis_interaction_count_duration_until_first_guidance} shows the distribution of two metrics - time taken (in minutes) and number of interactions performed - before requesting guidance for the first time.

\paragraphHeadingSpace\noindent\colorbox{lightergray}{\textbf{Metric:} Time taken.} 
Analyzing for the shortest time, \group{} finds the most eagerness to request guidance~(median=0.28 minutes into the task) followed by \guidance{}~(0.45), \expert{}~(2.19), and \ai{}~(2.81).

\paragraphHeadingSpace\noindent\colorbox{lightergray}{\textbf{Metric:} Interactions performed.}
Analyzing for the least number of interactions, \group{} sees the most eagerness to request guidance~(median=0 interactions into the task) followed by \guidance{}~(2), \expert{}~(22), and \ai{}~(24). The high eagerness to access guidance under \group{} may indicate a belief that the stored knowledge from this source is more psychologically accessible~\cite{higgins1996activation}, given that the source comprises analysts who are peers and thus at the same level of hierarchy as users, unlike \ai{} and \expert{} who are perceived at a higher level. This aspect of \textit{accessibility} along with salience is central to users' knowledge activation~\cite{higgins1996activation}, and is well recognized in information processing in psychology.

\subsubsection{Behavioral Decision Making about Attributes}

\begin{figure}[H]
    \includegraphics[width=0.9\linewidth]{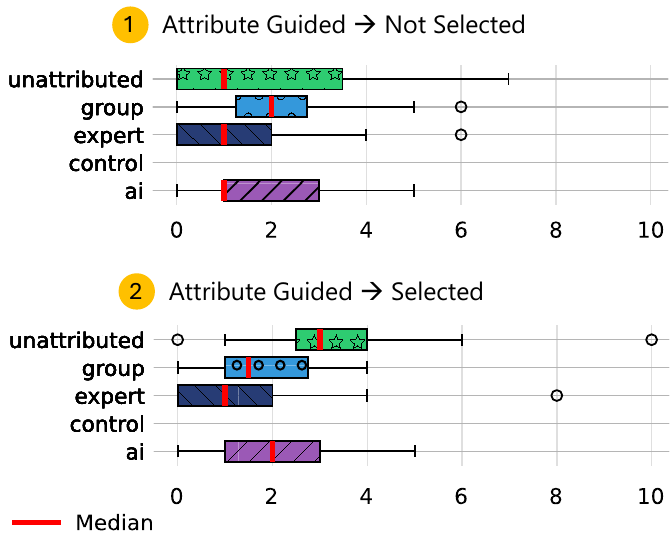}
    \centering
    \caption{Distribution of the number of times participants (1) did not select a guided attribute and (2) did select a guided attribute.}
    \label{fig:log_analysis_guidance_reliance}
    \Description{Distribution of the number of times participants (1) did not select a guided attribute and (2) did select a guided attribute.}
\end{figure}

Given our conceptualization of guidance as behavioral decision making about attributes, we analyzed how participants' interaction behavior with respect to attribute selection and deselection changed during the task relative to the received guidance (Figure~\ref{fig:log_analysis_guidance_reliance}).

\begin{enumerate}[left=0pt]
    
    \item \colorbox{lightergray}{[Attribute Guided $\rightarrow$ Not Selected]}
    \emph{Users received an attribute as guidance and never selected it.}
    \group{} (median=2.0) exhibited this behavior more than \ai{}, \expert{}, and \guidance{} (median=1.0). This directionally points to users' higher \textit{disagreement} with guidance coming from \group{} than from \ai{} or \expert{}, suggesting more benefit from latter among attributed sources. This metric is similar to Lu~et~al.'s ``disagreement'' metric~\cite{lu2021human}.

    \item \colorbox{lightergray}{[Attribute Guided $\rightarrow$ Selected]}
    \emph{Users received an attribute as guidance and selected it.}
    \guidance{} (median=3.0) exhibited this behavior more than \ai{} (median=2.0) than \group{} (median=1.5) than \expert{} (median=1.0). This measure of \textit{agreement} with guidance, shows no directional advantage among any of the attributed sources. 
    This metric is similar to Lu~et~al.'s ``agreement'' metric~\cite{lu2021human}.
\end{enumerate}

\subsection{Analysis of Stated Responses}
To complement the objective measures for utilization, we report findings from analyzing the subjective response(s) in \textbf{Pre-Study Questionnaire 2} and \textbf{Post-Study Questionnaire}. 
For detailed statistics, refer to Table~\ref{tab:questionnaire-scores} in Appendix~\ref{section:appendix-questionnaires}.

\begin{figure*}[!ht]
    \includegraphics[width=\linewidth]{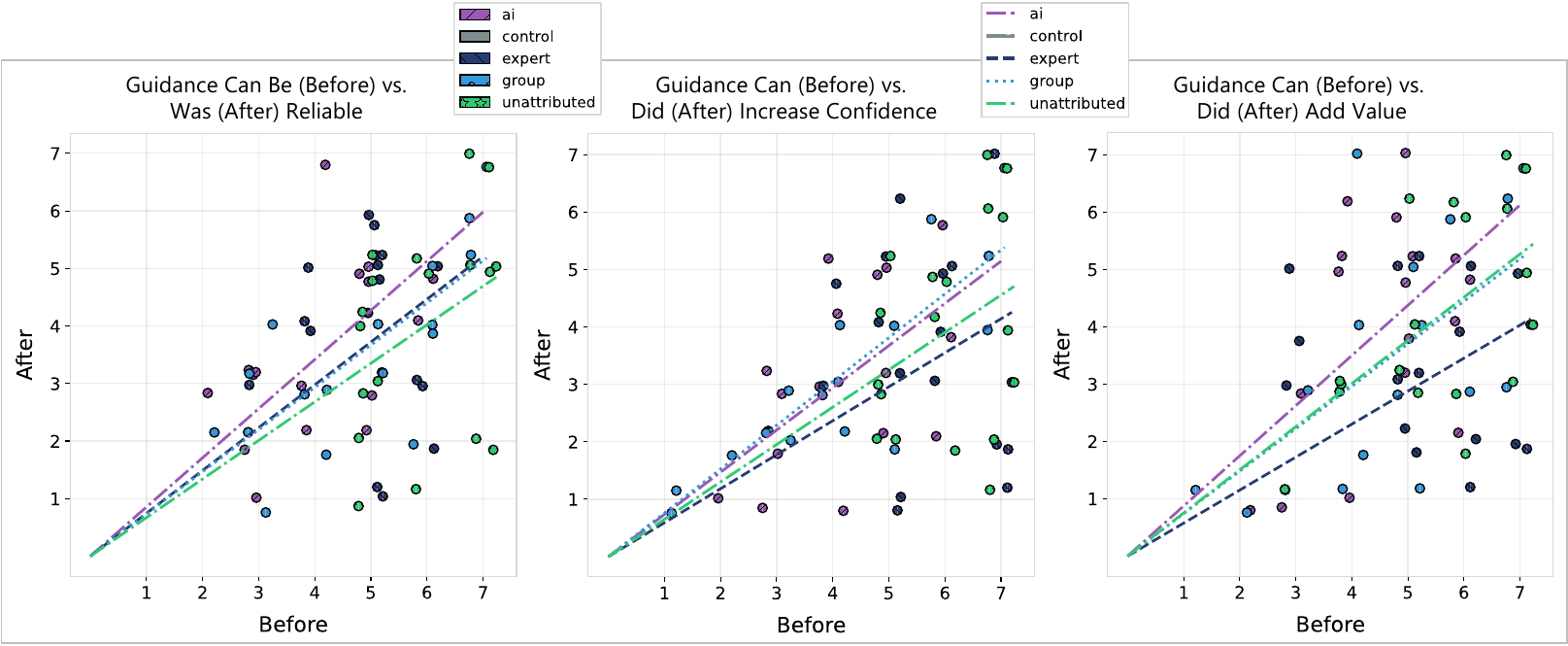}
    \centering
    \caption{Participants' opinion of guidance (``Reliability'', ``Confidence'', ``Value Add'') before and after the task, on a scale from 1 (Disagree) to 7 (Agree), as reported in the \textbf{Pre-Study Questionnaire 2} and \textbf{Post-Study Questionnaire}.~\control{} neither received guidance nor these questions. A slope--of the regression line--steeper (shallower) than the 45-degree line shows after-task responses higher (lower) than pre-task responses, per condition.}
    \label{fig:pre-post-study-guidance-opinion}
    \Description{Participants' opinion of guidance (``Reliability'', ``Confidence'', ``Value Add'') before and after the task.}
\end{figure*}

\subsubsection{Opinion about Guidance before and after analysis.} 
Figure~\ref{fig:pre-post-study-guidance-opinion} shows how participants' opinions about guidance (``Reliability'', ``Confidence'', ``Value Add'') changed after the task, on a scale from 1 (Disagree) to 7 (Agree), along with best fit regression lines. 
For each factor, we compute the difference between relevant participants' \emph{after} and \emph{before} scores.
Then, to test for statistical significance between the four conditions that received guidance (\ai{}, \expert{}, \group{}, \guidance{}), we utilize pairwise two-sided Mann Whitney U tests with Bonferroni correction.

\paragraphHeadingSpace\noindent\colorbox{lightergray}{\textbf{Metric:} Reliability.} \guidance{} (median=-2.0) scored lowest followed by \group{} (median=-1.0), then \expert{} and \ai{} (median=0.0), with no statistically significant differences.

\paragraphHeadingSpace\noindent\colorbox{lightergray}{\textbf{Metric:} Confidence.}
\guidance{} and \expert{} (median=-2.0) scored lowest followed by \group{} and \ai{} (median=-1.0), with no statistically significant differences.

\paragraphHeadingSpace\noindent\colorbox{lightergray}{\textbf{Metric:} Value Add.} 
\expert{} (median=-2.0) scored lowest followed by \group{}, \ai{}, and \guidance{} (median=-1.0), with no statistically significant differences.

\subsubsection{Reliance on- and regret using guidance.} 
\begin{figure*}[!ht]
    \includegraphics[width=0.8\linewidth]{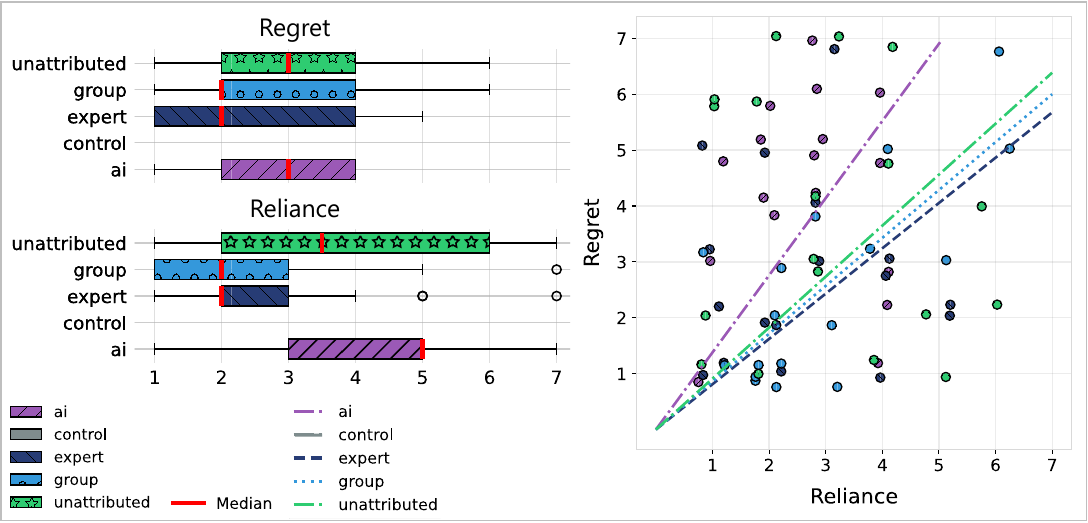}
    \centering
    \caption{Participants' response scores about their reliance on and regret using guidance, on a scale from 1 (None at all) to 7 (A lot), as reported in the \textbf{Post-Study Questionnaire}.~\control{} neither received guidance nor these questions. In the scatterplot, a slope--of the regression line--steeper (shallower) than the 45-degree line shows regret higher (lower) than reliance, per condition.}
    \label{fig:post-study-regret-reliance}
    \Description{}
\end{figure*}

Figure~\ref{fig:post-study-regret-reliance} shows users' response scores, on a scale from 1 (None at all) to 7 (A lot), on how much they relied on (``Reliance'') and regretted relying on (``Regret'') guidance during the task.

\paragraphHeadingSpace\noindent\colorbox{lightergray}{\textbf{Metric:} Reliance.}~\ai{} and \guidance{} reported higher reliance on guidance (median=3.0) than both \expert{} and \group{} (median=2.0).

\paragraphHeadingSpace\noindent\colorbox{lightergray}{\textbf{Metric:} Regret.}~\ai{} reported higher regret relying on guidance (median=5.0) than both \expert{} and \group{} (median=2.0), with \guidance{} in between (median=3.5).

\subsubsection{Regret in Guidance: A Deep-dive}


\begin{table*}[!ht]
   \caption{Two regression analyses focused on the question on \textbf{Regret} in the \textbf{Post-Study Questionnaire}. Features comprise stated responses, which are numerical variables, and attributed sources that comprise categorical variable and are one-hot encoded. \underline{\textbf{Bold and underline}} indicate statistically significant Pr>|t| at $\alpha = 0.05$.}
    \label{tab:regret-regression-ana}
    \centering
    \renewcommand{\arraystretch}{1.0} 
    \begin{tabular}{l|r|r}
        \toprule
        \textbf{Feature} & \textbf{Parameter Estimate} & \textbf{$Pr > |t|$} \\
        \hline
        \multicolumn{3}{c}{\cellcolor{lightergray}Analysis I} \\
        \hline
        \multicolumn{3}{c}{Regression Analysis: Target - Post study Regret. Features - Post study responses and attributed-Sources. ${Adj-R}^2=0.29$} \\
        \hline
        Guidance Was \textbf{Trustworthy} & 0.08 & 0.86 \\
        Guidance Came From A \textbf{Knowledgeable} Source & 0.23 & 0.16 \\
        Guidance Was \textbf{Reliable} & 0.39 & 0.38 \\
        Guidance Increased \textbf{Confidence} & -0.63 & \underline{\textbf{0.03}} \\
        Guidance Was \textbf{Relevant} & -0.17 & 0.59 \\
        Guidance Helped Avoid Potential \textbf{Pitfalls} & 0.41 & 0.13 \\
        Guidance Gave Valuable \textbf{Suggestions} & 0.13 & 0.58 \\
        Guidance \textbf{Added Value} & 0.34 & 0.34 \\
        Guidance Was \textbf{Appropriate} & -0.09 & 0.63 \\
        \ai{} & 1.46 & \underline{\textbf{0.02}} \\
        \expert{} & 0.65 & 0.28 \\
        Intercept (\group{}) & -0.03 & 0.97 \\
        \hline
        \multicolumn{3}{c}{\cellcolor{lightergray}Analysis II} \\
        \hline
        \multicolumn{3}{c}{Regression Analysis: Target - Post study Regret. Features - Post \textit{minus} Pre study responses and attributed-Sources.  ${Adj-R}^2=0.23$} \\
        \hline
        Guidance Was \textbf{Reliable} \textit{minus} Guidance can be \textbf{Reliable} & 0.38 & \underline{\textbf{0.04}} \\
        Guidance Increased \textbf{Confidence} \textit{minus} Guidance can \textbf{Increase Confidence} & -0.61 & \underline{\textbf{0.01}} \\
        Guidance \textbf{Added Value} \textit{minus} Guidance can \textbf{Add Value} & 0.40 & \underline{\textbf{0.04}} \\
        \ai{} & 1.06 & 0.08 \\
        \expert{} & -0.29 & 0.65 \\
        Intercept (\group{}) & 2.95 & \underline{\textbf{<0.0001}} \\
        \hline
    \end{tabular}
\end{table*}

In Section~\ref{section:introduction}, we introduced the threefold importance of regret for guidance systems: post-task regret may impact continued utilization of guidance, anticipatory regret may distort use of guidance~\cite{tzini2018role}, and managing expectations of users may be necessary for success. 
Taking a deeper dive, we performed a regression analysis of drivers of regret, summarized in
Table~\ref{tab:regret-regression-ana}. 
\textbf{Analysis I} finds that among the nine guidance features having post-task response, only ``Guidance Increased \textbf{Confidence}'' is considered an important driver of post-task \textbf{Regret}. The negative coefficient implies that as confidence increases, \textbf{Regret} decreases, which makes intuitive sense. Among sources, \ai{} is significant and its coefficient positive (relative to the baseline, \group{}, in this regression analysis), suggesting post-task \textbf{Regret} is higher for \ai{} than other two attributed sources, in the presence of these features. 

\textbf{Analysis II} examines effects of \emph{Expectation} on post-task Regret. In the Pre-Study Questionnaire 2, we collected responses on three guidance features, after priming about the source, but before the task was performed, giving measure of user's \emph{Expectation} on those three features. Then, in the post-task questionnaire (Post-Study Questionnaire), we collected responses on the same three features to give measures of \emph{Realization}. The \textit{change}--\emph{Realization} \textit{minus} \emph{Expectation}--shows significance for all three features. As ``Guidance Increased \textbf{Confidence}'' improves even more from \emph{Expectation}, \textbf{Regret} decreases further. Also, \group{} and \ai{} both contribute to higher \textbf{Regret} in the presence of these changes (noting that the coefficient for \ai{} is additive to the baseline, \group{}'s, coefficient). 

\subsection{Qualitative Feedback}

In the \textbf{Post-Study Questionnaire}, we asked participants to respond to four key questions related to the study task. We analyzed their responses by applying open coding~\cite{boyatzis1998transforming}, specifically, constant comparison and theoretical sampling~\cite{strauss1998basics}. Below, we list these four questions along with corresponding key takeaways.

\bpstart{\emph{``Describe your analysis strategy to identify relevant attributes.''}}  
Participants employed a mix of personal judgment and guidance. Some sought guidance early, integrating the recommendations with their own analysis, while others relied more on self-judgment initially. Many focused on attributes that provided clear insights into customer behavior, sales patterns, and return on investment, prioritizing those with distinct distributions, while avoiding more complex or less relevant attributes like ``IP address''.

\paragraphHeadingSpace\bpstart{\emph{``Describe your strategy with respect to designing the visualizations.''}}
Participants predominantly created visualizations emphasizing market demographics and sales drivers, as was required by the task.
Their primary strategy involved trial and error, with participants selecting relevant attributes, testing various combinations to identify patterns, and then refining the visualizations based on the data's ability to convey meaningful insights quickly and intuitively.
Most participants prioritized simplicity for managerial discussions, using scatter plots and bar charts while limiting encoded attributes to three for clarity.
Another participant strategy focused on creating visualizations that highlighted correlations. They experimented with different visualization types and configurations, such as line charts for time series data and scatter plots for comparing numeric variables.

\paragraphHeadingSpace\bpstart{\emph{``Do you have additional comments about the guidance you \textbf{received} during the task?''}}  
Feedback on the guidance received during the task was mixed. 
Some participants noted issues with the guidance, such as receiving attributes that were confusing or irrelevant to their analysis. For example, one participant found the initial attribute suggestions unhelpful due to a lack of clarity and connection between them, while another felt that specific attributes like ``ID''s were not useful for their aggregate-focused task. A few participants expressed frustration with the guidance's lack of explanations and rationale, which hindered their understanding and trust in the suggestions.
Conversely, some participants found the guidance beneficial, appreciating its ability to offer new attribute ideas or confirming their own choices. They also noted that explanations or additional context would have enhanced its utility, particularly when attributes were suggested one at a time.

\paragraphHeadingSpace\bpstart{\emph{``Do you have additional comments about the guidance you \textbf{may receive} from different sources shown during the task?''}}
Participants expressed a slight preference for human guidance over AI, citing the ability of humans to provide more contextual and tailored advice. One participant suggested that combining AI with human expert guidance would be more effective. Others noted that internal guidance from colleagues might be more relevant due to familiarity with organizational needs, with one remarking, \emph{``I struggle to decide how I feel about the group of analysts vs the human expert. Of course, I am inclined to believe/take guidance from an expert in [visualization] and/or data analytics, but I may lean heavier on a `less experienced' group (or single) person in the company as they are aware how the business runs, what is important to the executives, and if that one director of marketing *really* loves pie charts (like a lot). An expert in the field can certainly provide great feedback on best practices, new methods, and what the research shows, but they cannot comment on the idiosyncrasies of the company like internal analysts could.''}

\subsection{Overall Task Fidelity and System Usability}

\aptLtoX{\begin{figure*}[!ht]
        \centering
        \includegraphics[width=\columnwidth]{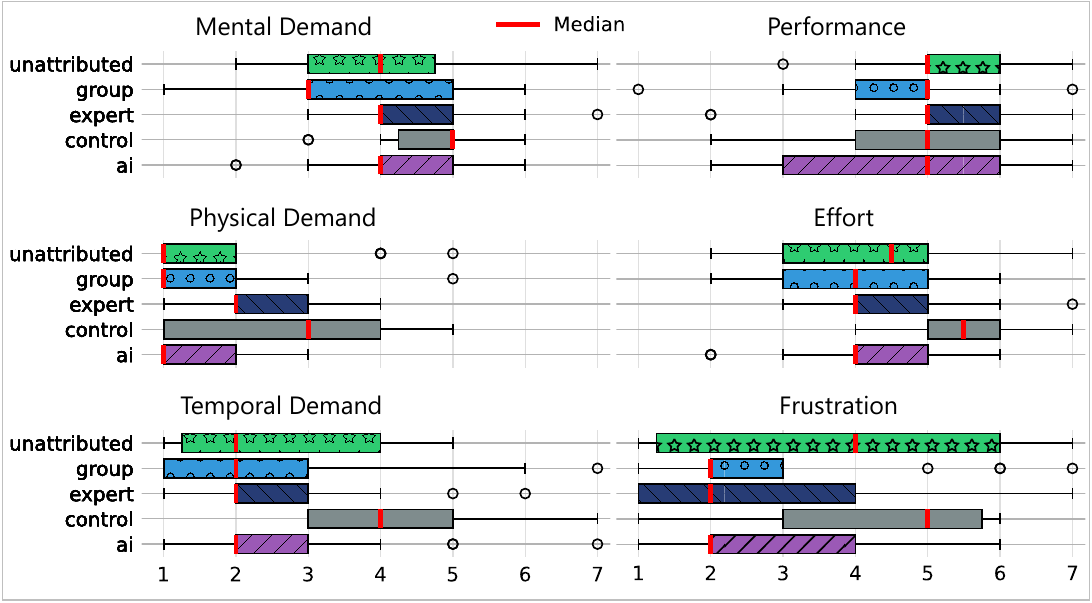}
        \caption{Participants' self-reported responses about the overall fidelity of the task on a scale from 1 (Low) to 7 (High), as reported in the \textbf{Post-Study Questionnaire}.}
        \Description{Participants' self-reported responses about the overall fidelity of the task on a scale from 1 (Low) to 7 (High), as reported in the \textbf{Post-Study Questionnaire}.}
        \label{fig:task-fidelity}
\end{figure*}
\begin{table}
        \centering
\begin{tabular}{l|c|c|c|c|c}
    \toprule
    \cellcolor{white}\textbf{Metric} & {\textcolor{colorai}{\textbf{AI}}} & {\textcolor{colorexpert}{\textbf{Expert}}} & {\textcolor{colorgroup}{\textbf{Group}}} & \textcolor{colorguidance}{\textbf{Unattributed}} & {\textcolor{colorcontrol}{\textbf{Control}}} \\
    \midrule
    Mental Demand & {\textcolor{colorai}{4.0}} & {\textcolor{colorexpert}{4.0}} & \textcolor{colorgroup}{3.0} & \textcolor{colorguidance}{4.0} & \textcolor{colorcontrol}{\underline{\textbf{5.0}}} \\
    Physical Demand & \textcolor{colorai}{1.0} & \textcolor{colorexpert}{2.0} & \textcolor{colorgroup}{1.0} & \textcolor{colorguidance}{1.0} & \textcolor{colorcontrol}{\underline{\textbf{3.0}}} \\
    Temporal Demand & \textcolor{colorai}{2.0} & \textcolor{colorexpert}{2.0} & \textcolor{colorgroup}{2.0} & \textcolor{colorguidance}{2.0} & \textcolor{colorcontrol}{\underline{\textbf{4.0}}} \\
    Performance & \textcolor{colorai}{\underline{\textbf{5.0}}} & \textcolor{colorexpert}{\underline{\textbf{5.0}}} & \textcolor{colorgroup}{\underline{\textbf{5.0}}} & \textcolor{colorguidance}{\underline{\textbf{5.0}}} & \textcolor{colorcontrol}{\underline{\textbf{5.0}}} \\
    Effort & \textcolor{colorai}{4.0} & \textcolor{colorexpert}{4.0} & \textcolor{colorgroup}{4.0} & \textcolor{colorguidance}{4.5} & \textcolor{colorcontrol}{\underline{\textbf{5.5}}} \\
    Frustration & \textcolor{colorai}{2.0} & \textcolor{colorexpert}{2.0} & \textcolor{colorgroup}{2.0} & \textcolor{colorguidance}{4.0} & \textcolor{colorcontrol}{\underline{\textbf{5.0}}} \\
    \bottomrule
\end{tabular}
\caption{Median scores about the fidelity of the task, as self-reported by participants on a scale from 1 (Low) to 7 (High), in the \textbf{Post-Study Questionnaire}. \underline{\textbf{Bold and underline}} indicate the largest values in each row.}
        \label{tab:task-fidelity}
\end{table}}{
    
    \begin{figure*}[!ht]
        \centering
        \includegraphics[width=0.75\textwidth]{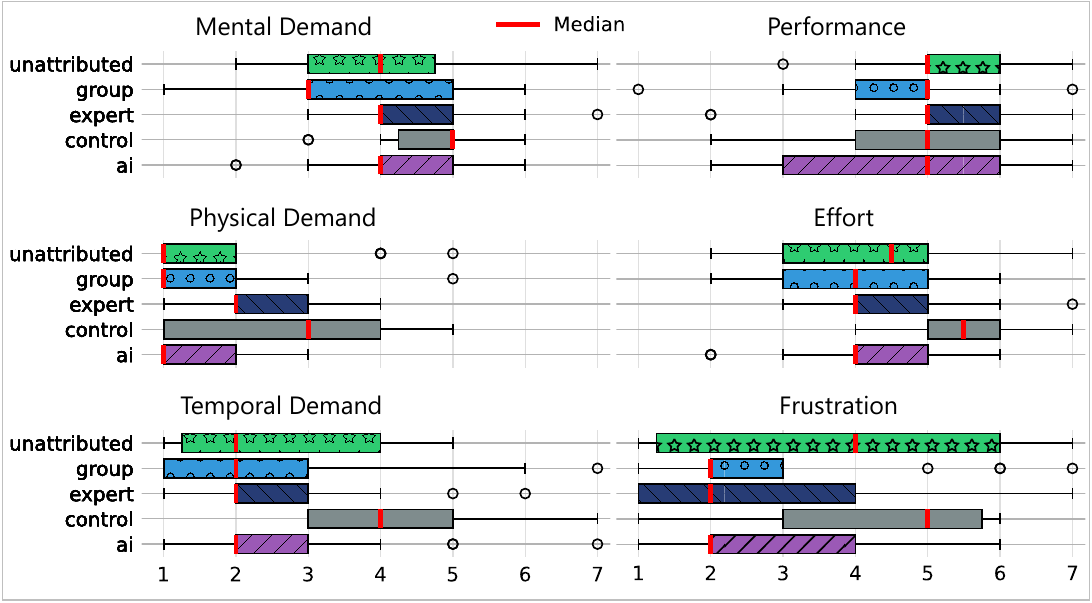}
        \caption{Participants' self-reported responses about the overall fidelity of the task on a scale from 1 (Low) to 7 (High), as reported in the \textbf{Post-Study Questionnaire}.}
        \Description{Participants' self-reported responses about the overall fidelity of the task on a scale from 1 (Low) to 7 (High), as reported in the \textbf{Post-Study Questionnaire}.}
        \label{fig:task-fidelity}
    \end{figure*}

    \begin{table}
        \setlength{\tabcolsep}{0.25em}
        \centering
    \begin{tabular}{l|>{\color{colorai}}c|>{\color{colorexpert}}c|>{\color{colorgroup}}c|>{\color{colorguidance}}c|>{\color{colorcontrol}}c}
    \toprule
    \cellcolor{white}\textbf{Metric} & \textbf{\ai{}} & \textbf{\expert{}} & \textbf{\group{}} & \textbf{\guidance{}} & \textbf{\control{}} \\
    \midrule
    Mental Demand & 4.0 & 4.0 & 3.0 & 4.0 & \underline{\textbf{5.0}} \\
    Physical Demand & 1.0 & 2.0 & 1.0 & 1.0 & \underline{\textbf{3.0}} \\
    Temporal Demand & 2.0 & 2.0 & 2.0 & 2.0 & \underline{\textbf{4.0}} \\
    Performance & \underline{\textbf{5.0}} & \underline{\textbf{5.0}} & \underline{\textbf{5.0}} & \underline{\textbf{5.0}} & \underline{\textbf{5.0}} \\
    Effort & 4.0 & 4.0 & 4.0 & 4.5 & \underline{\textbf{5.5}} \\
    Frustration & 2.0 & 2.0 & 2.0 & 4.0 & \underline{\textbf{5.0}} \\
    \bottomrule
    \end{tabular}
    \caption{Median scores about the fidelity of the task, as self-reported by participants on a scale from 1 (Low) to 7 (High), in the \textbf{Post-Study Questionnaire}. \underline{\textbf{Bold and underline}} indicate the largest values in each row.}
    \label{tab:task-fidelity}
    \end{table}
}

\paragraphHeadingSpace\noindent\colorbox{lightergray}{Task Fidelity.} Figure~\ref{fig:task-fidelity} and Table~\ref{tab:task-fidelity} show summary statistics of participant responses to questions in the \textbf{Post-Study Questionnaire} about the fidelity of the task.
\control{} reported the highest amount of mental demand, physical demand, temporal demand, effort, and frustration. All study conditions scored equally on performance.

\paragraphHeadingSpace\noindent\colorbox{lightergray}{System Usability.} Participants evaluated our study prototype using the System Usability Scale (SUS)~\cite{brooke1996sus}, reporting an overall mean score of 72.67. For individual study conditions, participants reported mean scores of 76.76~(\ai{}), 74.85~(\expert{}), 69.41~(\group{}), 74.31~(\guidance{}), and 68.19~(\control{}).


\section{Discussion}
\label{section:discussion}


In this section, we discuss how source-attribution in guidance affected users' usage and perception. For detailed statistics associated with participants scores to the two pre- and one post-study questionnaires, refer to Table~\ref{tab:questionnaire-scores} in Appendix~\ref{section:appendix-questionnaires}.

The nuances in how the perception of guidance changes with source-attribution, even when the quality of the guidance and the task remain identical, are captured in responses to \textbf{Pre-Study Questionnaire 2} and the \textbf{Post-Study Questionnaire}.
Recall that \textbf{Pre-Study Questionnaire 2} was administered after participants were primed about the respective source-condition, but before they started the task. Hence, questions in the \textbf{Pre-Study Questionnaire 2} captured ex-ante \textit{expectation} of participants, associated with the source. The \textbf{Post-Study Questionnaire} was administered after the completion of task, and thus represents post-task \textit{realization} of participants.

Comparing \ai{}, \expert{}, and \group{} across ``Pre Study Questionnaire 2: Questions about Guidance,'' we find that expectations about \textit{Reliable}, \textit{Increase-Confidence}, and \textit{Value-add} are similar across three source-conditions, except for \textit{Increase-Confidence}, where \expert{} scores higher. These scores indicate individual's baseline expectations about these characteristics of guidance. We compare participants' realizations after the task to see how much change occurred versus their expectations. Since survey responses are useful to interpret relative expectations instead of absolute ones, our approach provides clearer insights. 

Analyzing responses in the \textbf{Post-Study Questionnaire}, we find that \textit{Reliable} and \textit{Increase-Confidence} score similarly across three conditions; yet, \textit{Value-Add} scores higher for \ai{}. Digging deeper and comparing with expectations stated in the \textbf{Pre-Study Questionnaire 2}, in \textit{Value-Add}, post-task realization of participants meet pre-task expectations, only for \ai{}, but not for the two other sources, where their realizations fall short of expectations. In terms of \textit{Reliable} and \textit{Increase-Confidence}, realizations fall short of expectations for all three conditions. All this implies that in one out of three characteristics, realizations from \ai{} meets expectations, but for \expert{} and \group{} realizations fall short of expectations in all three characteristics. Now consider that our 1 (Disagree) - 7 (Agree) scale admits \textit{favorable} ratings only in the range 5 - 7. The post study questions find that, out of the 9 characteristics, \ai{} scores favorable in 3 of them, \expert{} in none, and \group{} in 1 characteristic. All this evidence suggests disposition toward \ai{} is no worse off than toward \expert{} or \group{}, and better in more characteristics, and should augur well for \ai{}.  

However, after task completion in the \textbf{Post-Study Questionnaire}, participants' response to the question asked of them  ``How Much [Did You] Regret Relying on Guidance,'' belie the above mentioned disposition toward \ai{}. Regret score is significantly higher for \ai{} than for \expert{} and \group{}. Yet, after task completion, participants acknowledge ``Guidance Came From A Knowledgeable Source'' and ``Guidance Gave Valuable Suggestions'' with scores for \ai{} at least as high as that of \expert{} and \group{}. Note that the weight of overall evidence points to something more nuanced than merely lower trust in \ai{} relative to other sources, which can account for resistance to adoption of guidance coming from \ai{}. We hypothesize that any enhanced task performance by a human with the help of \ai{} may create post-task dissonance about the ``messenger'' (being a non-human) although the ``message'' (quality of guidance) is the same across all sources. Future research can deep dive into dissonance to test this hypothesis and to find ways to mitigate.      

Moreover, our use of \guidance{} guidance provides a good source-agnostic benchmark for difference between realization and expectation. For \guidance{} guidance, in each of \textit{Reliable}, \textit{Increase-Confidence} and \textit{Value-add}, post-task scores are significantly lower than pre-study expectation, indicating that omission of source is undesirable as well. In addition, scores from ``Post Study Questionnaire: (\textit{What-If}) Questions about Guidance'' demonstrate that participants under \guidance{} guidance condition, assign a high score of $6$ to the counterfactual sources \expert{} and \group{}, which is higher than scores assigned to the counterfactual source \ai{}, for each of three characteristics: ``How Much \textbf{Faith} Would You Have,'' ``How Much Would You \textbf{Prefer},'' and ``How Much Would You \textbf{Rely}.'' These scores on post task counterfactual questions by users in \guidance{} source condition suggest two inferences: (i) an overall lower preference for \ai{}, and (ii) a desire for attributed-source, and (iii) that too pegged to humans - \expert{} or \group{}. 
Future research can dive deeper into these inferences.


\section{Limitations and Future Work}
\label{section:limitations}


Our study had some limitations. First, we did not provide explanations or justifications for the guidance, such as reasoning behind AI recommendations or context from human analysts, which could impact users’ trust and understanding. Moreover, \add{by design,} the task was somewhat subjective, as the relevance of attributes in data analysis can vary based on user interpretation, making it difficult to establish a clear ground truth. 
\add{Next, while our study was focused on attribute selection, future studies may explore other aspects of data preparation--such as collection, cleaning, transformation--and data analysis~\cite{gu2024data}.}
\add{Lastly, to help our participants select relevant attributes, we provided them with the attribute definitions, underlying data, summary visualizations, and the ability to create visualizations using one or more attributes with different aggregation levels applied. 
However, more involved user studies with other tools and features (e.g., advanced statistical indicators) are needed to understand how participants utilize the provided guidance to perform the task.}
Despite these limitations, our findings offer valuable insights into how guidance sources influence user behavior \add{during data-driven analysis and decision-making on subset selection}.

\add{We have also identified many opportunities as future work.} 
\add{First, in this study, we asked participants to self-report their ``Regret'' in utilizing the provided guidance on a Likert-scale ranging from 1 (None at all) to 7 (A lot); based on the reported scores, we found that participants in the \ai{} guidance condition expressed greater regret at the end of the task. Future work can specifically study the underlying factors contributing to participants' regret and also devise means to mitigate the same. This was beyond the scope of our experiment.}
\add{Second, examining how guidance influences user behavior over time and aligns with the sensemaking process may reveal when guidance is most useful.}
\add{Likewise, future work can also study when users may desire specific sources of guidance, e.g., \ai{} versus \expert{}. To study these temporal dynamics, a tool integrated with user mechanisms to switch between the different guidance sources is desirable.}
\add{Another promising direction for future work is examining how user expertise influences guidance utilization. For instance, do experts request guidance less frequently, or do they rely on it at specific stages of sensemaking? Investigating these behaviors could inform the design of adaptive guidance systems that tailor recommendations based on user expertise and their evolving needs.}
\edit{Lastly, through this study, we did} not offer specific prescriptions to improve \ai{} guidance systems, given \ai{}'s burgeoning role; however, findings from our study call for pointed future research devoted to dive deeper into the nuances of \ai{} as a source for guidance. Such research can invest in a large-scale guidance system, allowing more specific, actionable recommendations, but is outside the scope of this paper, and could be more prescriptive.


\section{Conclusion}
\label{section:conclusion}


We studied how the source of guidance -- \ai{}, \expert{}, or \group{} -- impacts users' perceptions and behaviors during an attribute selection task as part of data preparation. In a five-condition between-subjects study, including an \guidance{} guidance condition (i.e., guidance without any source attribution) and a no-guidance baseline (\control{}), study participants used a custom tool to select relevant attributes from an unfamiliar dataset, with guidance provided in the form of attribute suggestions. We controlled for guidance quality across conditions to ensure comparability. Our results showed that the source of guidance matters, as it influenced users' behavior in different ways during the task.
In particular, participants in the \ai{} guidance condition reported higher benefits but also expressing greater regret after the task. Overall, these findings strongly suggest the need to carefully consider guidance sources for effective guidance offerings in systems.


\begin{acks}
  This material is based upon work supported in part by NSF IIS-1750474 and a gift funding from Adobe, Inc. We are grateful to members of the Georgia Tech Visualization Lab, our user study participants, and anonymous reviewers for their constructive feedback at different stages of this work.
\end{acks}

\bibliographystyle{ACM-Reference-Format}




\begin{thebibliography}{101}


\ifx \showCODEN    \undefined \def \showCODEN     #1{\unskip}     \fi
\ifx \showISBNx    \undefined \def \showISBNx     #1{\unskip}     \fi
\ifx \showISBNxiii \undefined \def \showISBNxiii  #1{\unskip}     \fi
\ifx \showISSN     \undefined \def \showISSN      #1{\unskip}     \fi
\ifx \showLCCN     \undefined \def \showLCCN      #1{\unskip}     \fi
\ifx \shownote     \undefined \def \shownote      #1{#1}          \fi
\ifx \showarticletitle \undefined \def \showarticletitle #1{#1}   \fi
\ifx \showURL      \undefined \def \showURL       {\relax}        \fi
\providecommand\bibfield[2]{#2}
\providecommand\bibinfo[2]{#2}
\providecommand\natexlab[1]{#1}
\providecommand\showeprint[2][]{arXiv:#2}

\bibitem[big(2023)]%
        {bigeye}
 \bibinfo{year}{2023}\natexlab{}.
\newblock \bibinfo{title}{{Bigeye}}.
\newblock
\urldef\tempurl%
\url{https://www.bigeye.com/}
\showURL{%
Retrieved Jun 1, 2023 from \tempurl}


\bibitem[dat(2023)]%
        {datafold}
 \bibinfo{year}{2023}\natexlab{}.
\newblock \bibinfo{title}{{Datafold}}.
\newblock
\urldef\tempurl%
\url{https://www.datafold.com/}
\showURL{%
Retrieved Jun 1, 2023 from \tempurl}


\bibitem[mon(2023a)]%
        {monosi}
 \bibinfo{year}{2023}\natexlab{a}.
\newblock \bibinfo{title}{{Monosi}}.
\newblock
\urldef\tempurl%
\url{https://monosi.dev/}
\showURL{%
Retrieved Jun 1, 2023 from \tempurl}


\bibitem[mon(2023b)]%
        {montecarlodata}
 \bibinfo{year}{2023}\natexlab{b}.
\newblock \bibinfo{title}{{Monte Carlo Data}}.
\newblock
\urldef\tempurl%
\url{https://www.montecarlodata.com/}
\showURL{%
Retrieved Jun 1, 2023 from \tempurl}


\bibitem[sql(2023)]%
        {sqllineage}
 \bibinfo{year}{2023}\natexlab{}.
\newblock \bibinfo{title}{{SQLLineage}}.
\newblock
\urldef\tempurl%
\url{https://github.com/reata/sqllineage}
\showURL{%
Retrieved Jun 1, 2023 from \tempurl}


\bibitem[Banerjee(1992)]%
        {banerjee1992simple}
\bibfield{author}{\bibinfo{person}{Abhijit~V Banerjee}.}
  \bibinfo{year}{1992}\natexlab{}.
\newblock \showarticletitle{A simple model of herd behavior}.
\newblock \bibinfo{journal}{\emph{The quarterly journal of economics}}
  \bibinfo{volume}{107}, \bibinfo{number}{3} (\bibinfo{year}{1992}),
  \bibinfo{pages}{797--817}.
\newblock


\bibitem[Bansal et~al\mbox{.}(2019)]%
        {bansal2019case}
\bibfield{author}{\bibinfo{person}{Gagan Bansal}, \bibinfo{person}{Besmira
  Nushi}, \bibinfo{person}{Ece Kamar}, \bibinfo{person}{Dan Weld},
  \bibinfo{person}{Walter Lasecki}, {and} \bibinfo{person}{Eric Horvitz}.}
  \bibinfo{year}{2019}\natexlab{}.
\newblock \showarticletitle{A case for backward compatibility for human-ai
  teams}.
\newblock \bibinfo{journal}{\emph{arXiv preprint arXiv:1906.01148}}
  (\bibinfo{year}{2019}).
\newblock


\bibitem[Bell(1982)]%
        {bell1982regret}
\bibfield{author}{\bibinfo{person}{David~E Bell}.}
  \bibinfo{year}{1982}\natexlab{}.
\newblock \showarticletitle{Regret in decision making under uncertainty}.
\newblock \bibinfo{journal}{\emph{Operations research}} \bibinfo{volume}{30},
  \bibinfo{number}{5} (\bibinfo{year}{1982}), \bibinfo{pages}{961--981}.
\newblock


\bibitem[Borland et~al\mbox{.}(2019)]%
        {borland2019selection}
\bibfield{author}{\bibinfo{person}{David Borland}, \bibinfo{person}{Wenyuan
  Wang}, \bibinfo{person}{Jonathan Zhang}, \bibinfo{person}{Joshua Shrestha},
  {and} \bibinfo{person}{David Gotz}.} \bibinfo{year}{2019}\natexlab{}.
\newblock \showarticletitle{Selection bias tracking and detailed subset
  comparison for high-dimensional data}.
\newblock \bibinfo{journal}{\emph{IEEE Transactions on Visualization and
  Computer Graphics}} \bibinfo{volume}{26}, \bibinfo{number}{1}
  (\bibinfo{year}{2019}), \bibinfo{pages}{429--439}.
\newblock


\bibitem[Boyatzis(1998)]%
        {boyatzis1998transforming}
\bibfield{author}{\bibinfo{person}{Richard~E Boyatzis}.}
  \bibinfo{year}{1998}\natexlab{}.
\newblock \bibinfo{booktitle}{\emph{{Transforming Qualitative Information:
  Thematic Analysis and Code Development.}}}
\newblock \bibinfo{publisher}{{Sage Publications}}.
\newblock
\showISBNx{9780761909613}


\bibitem[Brackenbury et~al\mbox{.}(2018)]%
        {brackenbury2018draining}
\bibfield{author}{\bibinfo{person}{Will Brackenbury}, \bibinfo{person}{Rui
  Liu}, \bibinfo{person}{Mainack Mondal}, \bibinfo{person}{Aaron~J. Elmore},
  \bibinfo{person}{Blase Ur}, \bibinfo{person}{Kyle Chard}, {and}
  \bibinfo{person}{Michael~J. Franklin}.} \bibinfo{year}{2018}\natexlab{}.
\newblock \showarticletitle{{Draining the Data Swamp: A Similarity-Based
  Approach}}. In \bibinfo{booktitle}{\emph{Proceedings of the Workshop on
  Human-In-the-Loop Data Analytics}}. \bibinfo{publisher}{Association for
  Computing Machinery}, \bibinfo{address}{New York, NY, USA}, Article
  \bibinfo{articleno}{13}, \bibinfo{numpages}{7}~pages.
\newblock
\showISBNx{9781450358279}
\href{https://doi.org/10.1145/3209900.3209911}{doi:\nolinkurl{10.1145/3209900.3209911}}


\bibitem[Brooke et~al\mbox{.}(1996)]%
        {brooke1996sus}
\bibfield{author}{\bibinfo{person}{John Brooke} {et~al\mbox{.}}}
  \bibinfo{year}{1996}\natexlab{}.
\newblock \showarticletitle{SUS-A quick and dirty usability scale}.
\newblock \bibinfo{journal}{\emph{Usability evaluation in industry}}
  \bibinfo{volume}{189}, \bibinfo{number}{194} (\bibinfo{year}{1996}),
  \bibinfo{pages}{4--7}.
\newblock


\bibitem[Brynjolfsson and McElheran(2016)]%
        {brynjolfsson2016rapid}
\bibfield{author}{\bibinfo{person}{Erik Brynjolfsson} {and}
  \bibinfo{person}{Kristina McElheran}.} \bibinfo{year}{2016}\natexlab{}.
\newblock \showarticletitle{The rapid adoption of data-driven decision-making}.
\newblock \bibinfo{journal}{\emph{American Economic Review}}
  \bibinfo{volume}{106}, \bibinfo{number}{5} (\bibinfo{year}{2016}),
  \bibinfo{pages}{133--139}.
\newblock


\bibitem[Bu{\c{c}}inca et~al\mbox{.}(2021)]%
        {buccinca2021trust}
\bibfield{author}{\bibinfo{person}{Zana Bu{\c{c}}inca},
  \bibinfo{person}{Maja~Barbara Malaya}, {and} \bibinfo{person}{Krzysztof~Z
  Gajos}.} \bibinfo{year}{2021}\natexlab{}.
\newblock \showarticletitle{To trust or to think: cognitive forcing functions
  can reduce overreliance on AI in AI-assisted decision-making}.
\newblock \bibinfo{journal}{\emph{Proceedings of the ACM on Human-computer
  Interaction}} \bibinfo{volume}{5}, \bibinfo{number}{CSCW1}
  (\bibinfo{year}{2021}), \bibinfo{pages}{1--21}.
\newblock


\bibitem[Cafarella et~al\mbox{.}(2009)]%
        {cafarella2009data}
\bibfield{author}{\bibinfo{person}{Michael~J. Cafarella}, \bibinfo{person}{Alon
  Halevy}, {and} \bibinfo{person}{Nodira Khoussainova}.}
  \bibinfo{year}{2009}\natexlab{}.
\newblock \showarticletitle{{Data Integration for the Relational Web}}.
\newblock \bibinfo{journal}{\emph{Proceedings of the VLDB Endowment}}
  \bibinfo{volume}{2}, \bibinfo{number}{1} (\bibinfo{year}{2009}),
  \bibinfo{pages}{1090–1101}.
\newblock
\showISSN{2150-8097}
\href{https://doi.org/10.14778/1687627.1687750}{doi:\nolinkurl{10.14778/1687627.1687750}}


\bibitem[Ceneda et~al\mbox{.}(2020)]%
        {ceneda2020guide}
\bibfield{author}{\bibinfo{person}{Davide Ceneda}, \bibinfo{person}{Natalia
  Andrienko}, \bibinfo{person}{Gennady Andrienko}, \bibinfo{person}{Theresia
  Gschwandtner}, \bibinfo{person}{Silvia Miksch}, \bibinfo{person}{Nikolaus
  Piccolotto}, \bibinfo{person}{Tobias Schreck}, \bibinfo{person}{Marc Streit},
  \bibinfo{person}{Josef Suschnigg}, {and} \bibinfo{person}{Christian
  Tominski}.} \bibinfo{year}{2020}\natexlab{}.
\newblock \showarticletitle{Guide me in analysis: A framework for guidance
  designers}. In \bibinfo{booktitle}{\emph{Computer Graphics Forum}},
  Vol.~\bibinfo{volume}{39}. Wiley Online Library, \bibinfo{pages}{269--288}.
\newblock


\bibitem[Ceneda et~al\mbox{.}(2016)]%
        {ceneda2016characterizing}
\bibfield{author}{\bibinfo{person}{Davide Ceneda}, \bibinfo{person}{Theresia
  Gschwandtner}, \bibinfo{person}{Thorsten May}, \bibinfo{person}{Silvia
  Miksch}, \bibinfo{person}{Hans-J{\"o}rg Schulz}, \bibinfo{person}{Marc
  Streit}, {and} \bibinfo{person}{Christian Tominski}.}
  \bibinfo{year}{2016}\natexlab{}.
\newblock \showarticletitle{Characterizing guidance in visual analytics}.
\newblock \bibinfo{journal}{\emph{IEEE TVCG}} (\bibinfo{year}{2016}).
\newblock


\bibitem[Ceneda et~al\mbox{.}(2017)]%
        {ceneda2017amending}
\bibfield{author}{\bibinfo{person}{Davide Ceneda}, \bibinfo{person}{Theresia
  Gschwandtner}, \bibinfo{person}{Thorsten May}, \bibinfo{person}{Silvia
  Miksch}, \bibinfo{person}{Hans-J{\"o}rg Schulz}, \bibinfo{person}{Marc
  Streit}, {and} \bibinfo{person}{Christian Tominski}.}
  \bibinfo{year}{2017}\natexlab{}.
\newblock \showarticletitle{{Amending the Characterization of Guidance in
  Visual Analytics}}.
\newblock \bibinfo{journal}{\emph{arXiv preprint arXiv:1710.06615}}
  (\bibinfo{year}{2017}).
\newblock


\bibitem[Ceneda et~al\mbox{.}(2019)]%
        {ceneda2019review}
\bibfield{author}{\bibinfo{person}{Davide Ceneda}, \bibinfo{person}{Theresia
  Gschwandtner}, {and} \bibinfo{person}{Silvia Miksch}.}
  \bibinfo{year}{2019}\natexlab{}.
\newblock \showarticletitle{A review of guidance approaches in visual data
  analysis: A multifocal perspective}. In \bibinfo{booktitle}{\emph{Computer
  Graphics Forum}}, Vol.~\bibinfo{volume}{38}. Wiley Online Library,
  \bibinfo{pages}{861--879}.
\newblock


\bibitem[Chen(2008)]%
        {chen2008herd}
\bibfield{author}{\bibinfo{person}{Yi-Fen Chen}.}
  \bibinfo{year}{2008}\natexlab{}.
\newblock \showarticletitle{Herd behavior in purchasing books online}.
\newblock \bibinfo{journal}{\emph{Computers in Human Behavior}}
  \bibinfo{volume}{24}, \bibinfo{number}{5} (\bibinfo{year}{2008}),
  \bibinfo{pages}{1977--1992}.
\newblock


\bibitem[Chiang and Yin(2021)]%
        {chiang2021you}
\bibfield{author}{\bibinfo{person}{Chun-Wei Chiang} {and} \bibinfo{person}{Ming
  Yin}.} \bibinfo{year}{2021}\natexlab{}.
\newblock \showarticletitle{You’d better stop! Understanding human reliance
  on machine learning models under covariate shift}. In
  \bibinfo{booktitle}{\emph{Proceedings of the 13th ACM Web Science Conference
  2021}}. \bibinfo{pages}{120--129}.
\newblock


\bibitem[Chong et~al\mbox{.}(2022)]%
        {chong2022human}
\bibfield{author}{\bibinfo{person}{Leah Chong}, \bibinfo{person}{Guanglu
  Zhang}, \bibinfo{person}{Kosa Goucher-Lambert}, \bibinfo{person}{Kenneth
  Kotovsky}, {and} \bibinfo{person}{Jonathan Cagan}.}
  \bibinfo{year}{2022}\natexlab{}.
\newblock \showarticletitle{Human confidence in artificial intelligence and in
  themselves: The evolution and impact of confidence on adoption of AI advice}.
\newblock \bibinfo{journal}{\emph{Computers in Human Behavior}}
  \bibinfo{volume}{127} (\bibinfo{year}{2022}), \bibinfo{pages}{107018}.
\newblock


\bibitem[Collins et~al\mbox{.}(2018)]%
        {collins2018guidance}
\bibfield{author}{\bibinfo{person}{Christopher Collins},
  \bibinfo{person}{Natalia Andrienko}, \bibinfo{person}{Tobias Schreck},
  \bibinfo{person}{Jing Yang}, \bibinfo{person}{Jaegul Choo},
  \bibinfo{person}{Ulrich Engelke}, \bibinfo{person}{Amit Jena}, {and}
  \bibinfo{person}{Tim Dwyer}.} \bibinfo{year}{2018}\natexlab{}.
\newblock \showarticletitle{{Guidance in the human–machine analytics
  process}}.
\newblock \bibinfo{journal}{\emph{Visual Informatics}} \bibinfo{volume}{2},
  \bibinfo{number}{3} (\bibinfo{year}{2018}), \bibinfo{pages}{166--180}.
\newblock
\showISSN{2468-502X}
\href{https://doi.org/10.1016/j.visinf.2018.09.003}{doi:\nolinkurl{10.1016/j.visinf.2018.09.003}}


\bibitem[Deng et~al\mbox{.}(2017)]%
        {deng2017data}
\bibfield{author}{\bibinfo{person}{Dong Deng}, \bibinfo{person}{Raul~Castro
  Fernandez}, \bibinfo{person}{Ziawasch Abedjan}, \bibinfo{person}{Sibo Wang},
  \bibinfo{person}{Michael Stonebraker}, \bibinfo{person}{Ahmed~K Elmagarmid},
  \bibinfo{person}{Ihab~F Ilyas}, \bibinfo{person}{Samuel Madden},
  \bibinfo{person}{Mourad Ouzzani}, {and} \bibinfo{person}{Nan Tang}.}
  \bibinfo{year}{2017}\natexlab{}.
\newblock \showarticletitle{{The Data Civilizer System}}. In
  \bibinfo{booktitle}{\emph{CIDR}}.
\newblock
\urldef\tempurl%
\url{https://dblp.org/rec/conf/cidr/DengFAWSEIMO017.html}
\showURL{%
\tempurl}


\bibitem[Dix et~al\mbox{.}(2003)]%
        {dix2003human}
\bibfield{author}{\bibinfo{person}{Alan Dix}, \bibinfo{person}{Janet Finlay},
  \bibinfo{person}{Gregory~D Abowd}, {and} \bibinfo{person}{Russell Beale}.}
  \bibinfo{year}{2003}\natexlab{}.
\newblock \bibinfo{booktitle}{\emph{Human-computer interaction}}.
\newblock \bibinfo{publisher}{Pearson Education}.
\newblock


\bibitem[Duan et~al\mbox{.}(2009)]%
        {duan2009informational}
\bibfield{author}{\bibinfo{person}{Wenjing Duan}, \bibinfo{person}{Bin Gu},
  {and} \bibinfo{person}{Andrew~B Whinston}.} \bibinfo{year}{2009}\natexlab{}.
\newblock \showarticletitle{Informational cascades and software adoption on the
  internet: an empirical investigation}.
\newblock \bibinfo{journal}{\emph{MIS quarterly}} (\bibinfo{year}{2009}),
  \bibinfo{pages}{23--48}.
\newblock


\bibitem[Elish(2019)]%
        {elish2019moral}
\bibfield{author}{\bibinfo{person}{Madeleine~Clare Elish}.}
  \bibinfo{year}{2019}\natexlab{}.
\newblock \showarticletitle{Moral crumple zones: Cautionary tales in
  human-robot interaction (pre-print)}.
\newblock \bibinfo{journal}{\emph{Engaging Science, Technology, and Society
  (pre-print)}} (\bibinfo{year}{2019}).
\newblock


\bibitem[Endert et~al\mbox{.}(2012)]%
        {endert2012semanticinteraction}
\bibfield{author}{\bibinfo{person}{Alex Endert}, \bibinfo{person}{Patrick
  Fiaux}, {and} \bibinfo{person}{Chris North}.}
  \bibinfo{year}{2012}\natexlab{}.
\newblock \showarticletitle{Semantic Interaction for Visual Text Analytics}. In
  \bibinfo{booktitle}{\emph{Proceedings of the SIGCHI Conference on Human
  Factors in Computing Systems}} (Austin, Texas, USA)
  \emph{(\bibinfo{series}{CHI '12})}. \bibinfo{publisher}{Association for
  Computing Machinery}, \bibinfo{address}{New York, NY, USA},
  \bibinfo{pages}{473–482}.
\newblock
\showISBNx{9781450310154}
\href{https://doi.org/10.1145/2207676.2207741}{doi:\nolinkurl{10.1145/2207676.2207741}}


\bibitem[Engels(1996)]%
        {engels1996planning}
\bibfield{author}{\bibinfo{person}{Robert Engels}.}
  \bibinfo{year}{1996}\natexlab{}.
\newblock \showarticletitle{Planning tasks for Knowledge Discovery in
  Databases; Performing Task-Oriented User-Guidance.}. In
  \bibinfo{booktitle}{\emph{KDD}}. \bibinfo{pages}{170--175}.
\newblock


\bibitem[Engstr{\"o}m and Forsell(2018)]%
        {engstrom2018demand}
\bibfield{author}{\bibinfo{person}{Per Engstr{\"o}m} {and}
  \bibinfo{person}{Eskil Forsell}.} \bibinfo{year}{2018}\natexlab{}.
\newblock \showarticletitle{Demand effects of consumers’ stated and revealed
  preferences}.
\newblock \bibinfo{journal}{\emph{Journal of Economic Behavior \&
  Organization}}  \bibinfo{volume}{150} (\bibinfo{year}{2018}),
  \bibinfo{pages}{43--61}.
\newblock


\bibitem[Fehr et~al\mbox{.}(2005)]%
        {fehr2005neuroeconomic}
\bibfield{author}{\bibinfo{person}{Ernst Fehr}, \bibinfo{person}{Urs
  Fischbacher}, {and} \bibinfo{person}{Michael Kosfeld}.}
  \bibinfo{year}{2005}\natexlab{}.
\newblock \showarticletitle{Neuroeconomic foundations of trust and social
  preferences: initial evidence}.
\newblock \bibinfo{journal}{\emph{American Economic Review}}
  \bibinfo{volume}{95}, \bibinfo{number}{2} (\bibinfo{year}{2005}),
  \bibinfo{pages}{346--351}.
\newblock


\bibitem[Feltovich et~al\mbox{.}(2006)]%
        {feltovich2006studies}
\bibfield{author}{\bibinfo{person}{Paul~J Feltovich},
  \bibinfo{person}{Michael~J Prietula}, \bibinfo{person}{K~Anders Ericsson},
  {et~al\mbox{.}}} \bibinfo{year}{2006}\natexlab{}.
\newblock \showarticletitle{Studies of expertise from psychological
  perspectives}.
\newblock \bibinfo{journal}{\emph{The Cambridge handbook of expertise and
  expert performance}} (\bibinfo{year}{2006}), \bibinfo{pages}{41--67}.
\newblock


\bibitem[Fernandez et~al\mbox{.}(2018)]%
        {fernandez2018aurum}
\bibfield{author}{\bibinfo{person}{Raul~Castro Fernandez},
  \bibinfo{person}{Ziawasch Abedjan}, \bibinfo{person}{Famien Koko},
  \bibinfo{person}{Gina Yuan}, \bibinfo{person}{Samuel Madden}, {and}
  \bibinfo{person}{Michael Stonebraker}.} \bibinfo{year}{2018}\natexlab{}.
\newblock \showarticletitle{Aurum: A data discovery system}. In
  \bibinfo{booktitle}{\emph{2018 IEEE 34th International Conference on Data
  Engineering (ICDE)}}. IEEE.
\newblock


\bibitem[Fette and Melnikov(2011)]%
        {fette2011websocket}
\bibfield{author}{\bibinfo{person}{Ian Fette} {and} \bibinfo{person}{Alexey
  Melnikov}.} \bibinfo{year}{2011}\natexlab{}.
\newblock \bibinfo{booktitle}{\emph{The websocket protocol}}.
\newblock \bibinfo{type}{{T}echnical {R}eport}.
\newblock


\bibitem[Fishburn et~al\mbox{.}(1979)]%
        {fishburn1979utility}
\bibfield{author}{\bibinfo{person}{Peter~C Fishburn}, \bibinfo{person}{Peter~C
  Fishburn}, {et~al\mbox{.}}} \bibinfo{year}{1979}\natexlab{}.
\newblock \bibinfo{booktitle}{\emph{Utility theory for decision making}}.
\newblock \bibinfo{publisher}{Krieger NY}.
\newblock


\bibitem[Fodor(2002)]%
        {fodor2002survey}
\bibfield{author}{\bibinfo{person}{Imola~K Fodor}.}
  \bibinfo{year}{2002}\natexlab{}.
\newblock \bibinfo{booktitle}{\emph{A survey of dimension reduction
  techniques}}.
\newblock \bibinfo{type}{{T}echnical {R}eport}. \bibinfo{institution}{Lawrence
  Livermore National Lab., CA (US)}.
\newblock


\bibitem[Foundation(2024)]%
        {python}
\bibfield{author}{\bibinfo{person}{Python~Software Foundation}.}
  \bibinfo{year}{2024}\natexlab{}.
\newblock \bibinfo{title}{Python}.
\newblock
\urldef\tempurl%
\url{https://www.python.org/}
\showURL{%
Retrieved June 13, 2024 from \tempurl}


\bibitem[Gajos and Mamykina(2022)]%
        {gajos2022people}
\bibfield{author}{\bibinfo{person}{Krzysztof~Z Gajos} {and}
  \bibinfo{person}{Lena Mamykina}.} \bibinfo{year}{2022}\natexlab{}.
\newblock \showarticletitle{Do people engage cognitively with AI? Impact of AI
  assistance on incidental learning}. In \bibinfo{booktitle}{\emph{Proceedings
  of the 27th International Conference on Intelligent User Interfaces}}.
  \bibinfo{pages}{794--806}.
\newblock


\bibitem[Gaube et~al\mbox{.}(2021)]%
        {gaube2021ai}
\bibfield{author}{\bibinfo{person}{Susanne Gaube}, \bibinfo{person}{Harini
  Suresh}, \bibinfo{person}{Martina Raue}, \bibinfo{person}{Alexander Merritt},
  \bibinfo{person}{Seth~J Berkowitz}, \bibinfo{person}{Eva Lermer},
  \bibinfo{person}{Joseph~F Coughlin}, \bibinfo{person}{John~V Guttag},
  \bibinfo{person}{Errol Colak}, {and} \bibinfo{person}{Marzyeh Ghassemi}.}
  \bibinfo{year}{2021}\natexlab{}.
\newblock \showarticletitle{Do as AI say: susceptibility in deployment of
  clinical decision-aids}.
\newblock \bibinfo{journal}{\emph{NPJ digital medicine}} \bibinfo{volume}{4},
  \bibinfo{number}{1} (\bibinfo{year}{2021}), \bibinfo{pages}{31}.
\newblock


\bibitem[Gitelman(2013)]%
        {gitelman2013raw}
\bibfield{author}{\bibinfo{person}{Lisa Gitelman}.}
  \bibinfo{year}{2013}\natexlab{}.
\newblock \bibinfo{booktitle}{\emph{Raw data is an oxymoron}}.
\newblock \bibinfo{publisher}{MIT press}.
\newblock


\bibitem[Google(2024a)]%
        {angular}
\bibfield{author}{\bibinfo{person}{Google}.} \bibinfo{year}{2024}\natexlab{a}.
\newblock \bibinfo{title}{Angular}.
\newblock
\urldef\tempurl%
\url{https://angular.io/}
\showURL{%
Retrieved June 13, 2024 from \tempurl}


\bibitem[Google(2024b)]%
        {googlecloudlogging}
\bibfield{author}{\bibinfo{person}{Google}.} \bibinfo{year}{2024}\natexlab{b}.
\newblock \bibinfo{title}{Google Cloud Logging}.
\newblock
\urldef\tempurl%
\url{https://cloud.google.com/logging}
\showURL{%
Retrieved June 13, 2024 from \tempurl}


\bibitem[Gotz et~al\mbox{.}(2016)]%
        {gotz2016adaptive}
\bibfield{author}{\bibinfo{person}{David Gotz}, \bibinfo{person}{Shun Sun},
  {and} \bibinfo{person}{Nan Cao}.} \bibinfo{year}{2016}\natexlab{}.
\newblock \showarticletitle{Adaptive contextualization: Combating bias during
  high-dimensional visualization and data selection}. In
  \bibinfo{booktitle}{\emph{Proceedings of the 21st International Conference on
  Intelligent User Interfaces}}. \bibinfo{pages}{85--95}.
\newblock


\bibitem[Green(2022)]%
        {green2022flaws}
\bibfield{author}{\bibinfo{person}{Ben Green}.}
  \bibinfo{year}{2022}\natexlab{}.
\newblock \showarticletitle{The flaws of policies requiring human oversight of
  government algorithms}.
\newblock \bibinfo{journal}{\emph{Computer Law \& Security Review}}
  \bibinfo{volume}{45} (\bibinfo{year}{2022}), \bibinfo{pages}{105681}.
\newblock


\bibitem[Gu et~al\mbox{.}(2024)]%
        {gu2024data}
\bibfield{author}{\bibinfo{person}{Ken Gu}, \bibinfo{person}{Madeleine
  Grunde-McLaughlin}, \bibinfo{person}{Andrew McNutt}, \bibinfo{person}{Jeffrey
  Heer}, {and} \bibinfo{person}{Tim Althoff}.} \bibinfo{year}{2024}\natexlab{}.
\newblock \showarticletitle{How do data analysts respond to ai assistance? a
  wizard-of-oz study}. In \bibinfo{booktitle}{\emph{Proceedings of the CHI
  Conference on Human Factors in Computing Systems}}. \bibinfo{pages}{1--22}.
\newblock


\bibitem[Halevy et~al\mbox{.}(2016)]%
        {halevy2016goods}
\bibfield{author}{\bibinfo{person}{Alon Halevy}, \bibinfo{person}{Flip Korn},
  \bibinfo{person}{Natalya~F. Noy}, \bibinfo{person}{Christopher Olston},
  \bibinfo{person}{Neoklis Polyzotis}, \bibinfo{person}{Sudip Roy}, {and}
  \bibinfo{person}{Steven~Euijong Whang}.} \bibinfo{year}{2016}\natexlab{}.
\newblock \showarticletitle{{Goods: Organizing Google's Datasets}}. In
  \bibinfo{booktitle}{\emph{Proceedings of the 2016 International Conference on
  Management of Data}}. \bibinfo{publisher}{Association for Computing
  Machinery}, \bibinfo{address}{New York, NY, USA}, \bibinfo{pages}{795–806}.
\newblock
\showISBNx{9781450335317}
\href{https://doi.org/10.1145/2882903.2903730}{doi:\nolinkurl{10.1145/2882903.2903730}}


\bibitem[Ham(2013)]%
        {ham2013openrefine}
\bibfield{author}{\bibinfo{person}{Kelli Ham}.}
  \bibinfo{year}{2013}\natexlab{}.
\newblock \showarticletitle{OpenRefine (version 2.5). http://openrefine. org.
  Free, open-source tool for cleaning and transforming data}.
\newblock \bibinfo{journal}{\emph{Journal of the Medical Library Association:
  JMLA}} \bibinfo{volume}{101}, \bibinfo{number}{3} (\bibinfo{year}{2013}),
  \bibinfo{pages}{233}.
\newblock
\href{https://doi.org/10.3163/1536-5050.101.3.020}{doi:\nolinkurl{10.3163/1536-5050.101.3.020}}


\bibitem[Higgins(1996)]%
        {higgins1996activation}
\bibfield{author}{\bibinfo{person}{E~Tory Higgins}.}
  \bibinfo{year}{1996}\natexlab{}.
\newblock \showarticletitle{Activation: Accessibility, and salience}.
\newblock \bibinfo{journal}{\emph{Social psychology: Handbook of basic
  principles}} (\bibinfo{year}{1996}), \bibinfo{pages}{133--168}.
\newblock


\bibitem[Horvitz(1999)]%
        {horvitz1999principles}
\bibfield{author}{\bibinfo{person}{Eric Horvitz}.}
  \bibinfo{year}{1999}\natexlab{}.
\newblock \showarticletitle{{Principles of mixed-initiative user interfaces}}.
  In \bibinfo{booktitle}{\emph{{Proceedings of the SIGCHI conference on Human
  Factors in Computing Systems}}}. \bibinfo{pages}{159--166}.
\newblock


\bibitem[Jacobs et~al\mbox{.}(2021)]%
        {jacobs2021machine}
\bibfield{author}{\bibinfo{person}{Maia Jacobs}, \bibinfo{person}{Melanie~F
  Pradier}, \bibinfo{person}{Thomas~H McCoy~Jr}, \bibinfo{person}{Roy~H
  Perlis}, \bibinfo{person}{Finale Doshi-Velez}, {and}
  \bibinfo{person}{Krzysztof~Z Gajos}.} \bibinfo{year}{2021}\natexlab{}.
\newblock \showarticletitle{How machine-learning recommendations influence
  clinician treatment selections: the example of antidepressant selection}.
\newblock \bibinfo{journal}{\emph{Translational psychiatry}}
  \bibinfo{volume}{11}, \bibinfo{number}{1} (\bibinfo{year}{2021}),
  \bibinfo{pages}{108}.
\newblock


\bibitem[Jović et~al\mbox{.}(2015)]%
        {featureselection2015jovic}
\bibfield{author}{\bibinfo{person}{A. Jović}, \bibinfo{person}{K. Brkić},
  {and} \bibinfo{person}{N. Bogunović}.} \bibinfo{year}{2015}\natexlab{}.
\newblock \showarticletitle{A review of feature selection methods with
  applications}. In \bibinfo{booktitle}{\emph{2015 38th International
  Convention on Information and Communication Technology, Electronics and
  Microelectronics (MIPRO)}}. \bibinfo{pages}{1200--1205}.
\newblock
\href{https://doi.org/10.1109/MIPRO.2015.7160458}{doi:\nolinkurl{10.1109/MIPRO.2015.7160458}}


\bibitem[Kandel et~al\mbox{.}(2011)]%
        {kandel2011wrangler}
\bibfield{author}{\bibinfo{person}{Sean Kandel}, \bibinfo{person}{Andreas
  Paepcke}, \bibinfo{person}{Joseph Hellerstein}, {and}
  \bibinfo{person}{Jeffrey Heer}.} \bibinfo{year}{2011}\natexlab{}.
\newblock \showarticletitle{Wrangler: Interactive visual specification of data
  transformation scripts}. In \bibinfo{booktitle}{\emph{Proceedings of the
  SIGCHI Conference on Human Factors in Computing Systems}}.
  \bibinfo{pages}{3363--3372}.
\newblock


\bibitem[Kandel et~al\mbox{.}(2012)]%
        {kandel2012profiler}
\bibfield{author}{\bibinfo{person}{Sean Kandel}, \bibinfo{person}{Ravi Parikh},
  \bibinfo{person}{Andreas Paepcke}, \bibinfo{person}{Joseph~M. Hellerstein},
  {and} \bibinfo{person}{Jeffrey Heer}.} \bibinfo{year}{2012}\natexlab{}.
\newblock \showarticletitle{Profiler: Integrated Statistical Analysis and
  Visualization for Data Quality Assessment}. In
  \bibinfo{booktitle}{\emph{Proceedings of the International Working Conference
  on Advanced Visual Interfaces}} (Capri Island, Italy)
  \emph{(\bibinfo{series}{AVI '12})}. \bibinfo{publisher}{Association for
  Computing Machinery}, \bibinfo{address}{New York, NY, USA},
  \bibinfo{pages}{547–554}.
\newblock
\showISBNx{9781450312875}
\href{https://doi.org/10.1145/2254556.2254659}{doi:\nolinkurl{10.1145/2254556.2254659}}


\bibitem[Keim et~al\mbox{.}(2008)]%
        {keim2008visual}
\bibfield{author}{\bibinfo{person}{Daniel Keim}, \bibinfo{person}{Gennady
  Andrienko}, \bibinfo{person}{Jean-Daniel Fekete}, \bibinfo{person}{Carsten
  G{\"o}rg}, \bibinfo{person}{J{\"o}rn Kohlhammer}, {and} \bibinfo{person}{Guy
  Melan{\c{c}}on}.} \bibinfo{year}{2008}\natexlab{}.
\newblock \bibinfo{booktitle}{\emph{Visual analytics: Definition, process, and
  challenges}}.
\newblock \bibinfo{publisher}{Springer}.
\newblock


\bibitem[Kim et~al\mbox{.}(2023)]%
        {kim2023algorithms}
\bibfield{author}{\bibinfo{person}{Antino Kim}, \bibinfo{person}{Mochen Yang},
  {and} \bibinfo{person}{Jingjing Zhang}.} \bibinfo{year}{2023}\natexlab{}.
\newblock \showarticletitle{When algorithms err: Differential impact of early
  vs. late errors on users’ reliance on algorithms}.
\newblock \bibinfo{journal}{\emph{ACM Transactions on Computer-Human
  Interaction}} \bibinfo{volume}{30}, \bibinfo{number}{1}
  (\bibinfo{year}{2023}), \bibinfo{pages}{1--36}.
\newblock


\bibitem[Koulu(2020)]%
        {koulu2020human}
\bibfield{author}{\bibinfo{person}{Riikka Koulu}.}
  \bibinfo{year}{2020}\natexlab{}.
\newblock \showarticletitle{Human control over automation: EU policy and AI
  ethics}.
\newblock \bibinfo{journal}{\emph{Eur. J. Legal Stud.}}  \bibinfo{volume}{12}
  (\bibinfo{year}{2020}), \bibinfo{pages}{9}.
\newblock


\bibitem[Lai(2014)]%
        {lai2014expert}
\bibfield{author}{\bibinfo{person}{Ernest~K Lai}.}
  \bibinfo{year}{2014}\natexlab{}.
\newblock \showarticletitle{Expert advice for amateurs}.
\newblock \bibinfo{journal}{\emph{Journal of Economic Behavior \&
  Organization}}  \bibinfo{volume}{103} (\bibinfo{year}{2014}),
  \bibinfo{pages}{1--16}.
\newblock


\bibitem[L{\"a}pple and Barham(2019)]%
        {lapple2019learning}
\bibfield{author}{\bibinfo{person}{Doris L{\"a}pple} {and}
  \bibinfo{person}{Bradford~L Barham}.} \bibinfo{year}{2019}\natexlab{}.
\newblock \showarticletitle{How do learning ability, advice from experts and
  peers shape decision making?}
\newblock \bibinfo{journal}{\emph{Journal of Behavioral and Experimental
  Economics}}  \bibinfo{volume}{80} (\bibinfo{year}{2019}),
  \bibinfo{pages}{92--107}.
\newblock


\bibitem[Li et~al\mbox{.}(2017)]%
        {li2017feature}
\bibfield{author}{\bibinfo{person}{Jundong Li}, \bibinfo{person}{Kewei Cheng},
  \bibinfo{person}{Suhang Wang}, \bibinfo{person}{Fred Morstatter},
  \bibinfo{person}{Robert~P Trevino}, \bibinfo{person}{Jiliang Tang}, {and}
  \bibinfo{person}{Huan Liu}.} \bibinfo{year}{2017}\natexlab{}.
\newblock \showarticletitle{Feature selection: A data perspective}.
\newblock \bibinfo{journal}{\emph{ACM computing surveys (CSUR)}}
  \bibinfo{volume}{50}, \bibinfo{number}{6} (\bibinfo{year}{2017}),
  \bibinfo{pages}{1--45}.
\newblock


\bibitem[Lin et~al\mbox{.}(2006)]%
        {lin2006multiple}
\bibfield{author}{\bibinfo{person}{Chien-Huang Lin}, \bibinfo{person}{Wen-Hsien
  Huang}, {and} \bibinfo{person}{Marcel Zeelenberg}.}
  \bibinfo{year}{2006}\natexlab{}.
\newblock \showarticletitle{Multiple reference points in investor regret}.
\newblock \bibinfo{journal}{\emph{Journal of Economic Psychology}}
  \bibinfo{volume}{27}, \bibinfo{number}{6} (\bibinfo{year}{2006}),
  \bibinfo{pages}{781--792}.
\newblock


\bibitem[Liu et~al\mbox{.}(2022)]%
        {liu2022will}
\bibfield{author}{\bibinfo{person}{Yihe Liu}, \bibinfo{person}{Anushk Mittal},
  \bibinfo{person}{Diyi Yang}, {and} \bibinfo{person}{Amy Bruckman}.}
  \bibinfo{year}{2022}\natexlab{}.
\newblock \showarticletitle{Will AI console me when I lose my pet?
  Understanding perceptions of AI-mediated email writing}. In
  \bibinfo{booktitle}{\emph{Proceedings of the 2022 CHI conference on human
  factors in computing systems}}. \bibinfo{pages}{1--13}.
\newblock


\bibitem[Logg et~al\mbox{.}(2019)]%
        {logg2019algorithm}
\bibfield{author}{\bibinfo{person}{Jennifer~M Logg}, \bibinfo{person}{Julia~A
  Minson}, {and} \bibinfo{person}{Don~A Moore}.}
  \bibinfo{year}{2019}\natexlab{}.
\newblock \showarticletitle{Algorithm appreciation: People prefer algorithmic
  to human judgment}.
\newblock \bibinfo{journal}{\emph{Organizational Behavior and Human Decision
  Processes}}  \bibinfo{volume}{151} (\bibinfo{year}{2019}),
  \bibinfo{pages}{90--103}.
\newblock


\bibitem[Loomes and Sugden(1982)]%
        {loomes1982regret}
\bibfield{author}{\bibinfo{person}{Graham Loomes} {and} \bibinfo{person}{Robert
  Sugden}.} \bibinfo{year}{1982}\natexlab{}.
\newblock \showarticletitle{Regret theory: An alternative theory of rational
  choice under uncertainty}.
\newblock \bibinfo{journal}{\emph{The economic journal}} \bibinfo{volume}{92},
  \bibinfo{number}{368} (\bibinfo{year}{1982}), \bibinfo{pages}{805--824}.
\newblock


\bibitem[Lu and Yin(2021)]%
        {lu2021human}
\bibfield{author}{\bibinfo{person}{Zhuoran Lu} {and} \bibinfo{person}{Ming
  Yin}.} \bibinfo{year}{2021}\natexlab{}.
\newblock \showarticletitle{Human reliance on machine learning models when
  performance feedback is limited: Heuristics and risks}. In
  \bibinfo{booktitle}{\emph{Proceedings of the 2021 CHI Conference on Human
  Factors in Computing Systems}}. \bibinfo{pages}{1--16}.
\newblock


\bibitem[Mangin and Stoelinga(2011)]%
        {mangin2011peer}
\bibfield{author}{\bibinfo{person}{Melinda Mangin} {and} \bibinfo{person}{S
  Stoelinga}.} \bibinfo{year}{2011}\natexlab{}.
\newblock \showarticletitle{Peer? expert}.
\newblock \bibinfo{journal}{\emph{Journal of staff development}}
  \bibinfo{volume}{32}, \bibinfo{number}{3} (\bibinfo{year}{2011}),
  \bibinfo{pages}{48--52}.
\newblock


\bibitem[May et~al\mbox{.}(2011)]%
        {may2011guiding}
\bibfield{author}{\bibinfo{person}{Thorsten May}, \bibinfo{person}{Andreas
  Bannach}, \bibinfo{person}{James Davey}, \bibinfo{person}{Tobias Ruppert},
  {and} \bibinfo{person}{J{\"o}rn Kohlhammer}.}
  \bibinfo{year}{2011}\natexlab{}.
\newblock \showarticletitle{Guiding feature subset selection with an
  interactive visualization}. In \bibinfo{booktitle}{\emph{2011 IEEE Conference
  on Visual Analytics Science and Technology (VAST)}}. IEEE,
  \bibinfo{pages}{111--120}.
\newblock


\bibitem[Mellers et~al\mbox{.}(1998)]%
        {mellers1998judgment}
\bibfield{author}{\bibinfo{person}{Barbara~A Mellers}, \bibinfo{person}{Alan
  Schwartz}, {and} \bibinfo{person}{Alan~DJ Cooke}.}
  \bibinfo{year}{1998}\natexlab{}.
\newblock \showarticletitle{Judgment and decision making}.
\newblock \bibinfo{journal}{\emph{Annual review of psychology}}
  \bibinfo{volume}{49}, \bibinfo{number}{1} (\bibinfo{year}{1998}),
  \bibinfo{pages}{447--477}.
\newblock


\bibitem[Meservy et~al\mbox{.}(2021)]%
        {meservy2021searching}
\bibfield{author}{\bibinfo{person}{Tom Meservy}, \bibinfo{person}{Kelly~J
  Fadel}, \bibinfo{person}{Matthew~L Jensen}, {and} \bibinfo{person}{Michael
  Matthews}.} \bibinfo{year}{2021}\natexlab{}.
\newblock \showarticletitle{Searching for Expert or Peer Advice in Online
  Forums.}. In \bibinfo{booktitle}{\emph{AMCIS}}.
\newblock


\bibitem[Meshi et~al\mbox{.}(2012)]%
        {meshi2012expert}
\bibfield{author}{\bibinfo{person}{Dar Meshi}, \bibinfo{person}{Guido Biele},
  \bibinfo{person}{Christoph~W Korn}, {and} \bibinfo{person}{Hauke~R
  Heekeren}.} \bibinfo{year}{2012}\natexlab{}.
\newblock \showarticletitle{How expert advice influences decision making}.
\newblock \bibinfo{journal}{\emph{PLoS One}} \bibinfo{volume}{7},
  \bibinfo{number}{11} (\bibinfo{year}{2012}), \bibinfo{pages}{e49748}.
\newblock


\bibitem[Miller et~al\mbox{.}(2018)]%
        {miller2018making}
\bibfield{author}{\bibinfo{person}{Ren{\'e}e~J Miller},
  \bibinfo{person}{Fatemeh Nargesian}, \bibinfo{person}{Erkang Zhu},
  \bibinfo{person}{Christina Christodoulakis}, \bibinfo{person}{Ken~Q Pu},
  {and} \bibinfo{person}{Periklis Andritsos}.} \bibinfo{year}{2018}\natexlab{}.
\newblock \showarticletitle{{Making Open Data Transparent: Data Discovery on
  Open Data}}.
\newblock \bibinfo{journal}{\emph{IEEE Data Engineering Bulletin}}
  \bibinfo{volume}{41}, \bibinfo{number}{2} (\bibinfo{year}{2018}),
  \bibinfo{pages}{59--70}.
\newblock
\urldef\tempurl%
\url{http://sites.computer.org/debull/A18june/p59.pdf}
\showURL{%
\tempurl}


\bibitem[Moscati(2018)]%
        {moscati2018measuring}
\bibfield{author}{\bibinfo{person}{Ivan Moscati}.}
  \bibinfo{year}{2018}\natexlab{}.
\newblock \bibinfo{booktitle}{\emph{Measuring utility: From the marginal
  revolution to behavioral economics}}.
\newblock \bibinfo{publisher}{Oxford Studies in History of E}.
\newblock


\bibitem[Narechania et~al\mbox{.}(2023a)]%
        {narechania2023datacockpit}
\bibfield{author}{\bibinfo{person}{Arpit Narechania}, \bibinfo{person}{Surya
  Chakraborty}, \bibinfo{person}{Shivam Agarwal}, \bibinfo{person}{Atanu~R
  Sinha}, \bibinfo{person}{Ryan~A Rossi}, \bibinfo{person}{Fan Du},
  \bibinfo{person}{Jane Hoffswell}, \bibinfo{person}{Shunan Guo},
  \bibinfo{person}{Eunyee Koh}, \bibinfo{person}{Alex Endert}, {et~al\mbox{.}}}
  \bibinfo{year}{2023}\natexlab{a}.
\newblock \showarticletitle{DataCockpit: A Toolkit for Data Lake Navigation and
  Monitoring Utilizing Quality and Usage Information}. In
  \bibinfo{booktitle}{\emph{2023 IEEE International Conference on Big Data
  (BigData)}}. IEEE, \bibinfo{pages}{5305--5310}.
\newblock


\bibitem[Narechania et~al\mbox{.}(2022)]%
        {narechania2021lumos}
\bibfield{author}{\bibinfo{person}{Arpit Narechania}, \bibinfo{person}{Adam
  Coscia}, \bibinfo{person}{Emily Wall}, {and} \bibinfo{person}{Alex Endert}.}
  \bibinfo{year}{2022}\natexlab{}.
\newblock \showarticletitle{{Lumos: Increasing Awareness of Analytic Behavior
  during Visual Data Analysis}}.
\newblock \bibinfo{journal}{\emph{{IEEE Transactions on Visualization and
  Computer Graphics}}} \bibinfo{volume}{28}, \bibinfo{number}{1}
  (\bibinfo{year}{2022}), \bibinfo{pages}{1009--1018}.
\newblock
\href{https://doi.org/10.1109/TVCG.2021.3114827}{doi:\nolinkurl{10.1109/TVCG.2021.3114827}}


\bibitem[Narechania et~al\mbox{.}(2023b)]%
        {narechania2023datapilot}
\bibfield{author}{\bibinfo{person}{Arpit Narechania}, \bibinfo{person}{Fan Du},
  \bibinfo{person}{Atanu~R Sinha}, \bibinfo{person}{Ryan Rossi},
  \bibinfo{person}{Jane Hoffswell}, \bibinfo{person}{Shunan Guo},
  \bibinfo{person}{Eunyee Koh}, \bibinfo{person}{Shamkant~B Navathe}, {and}
  \bibinfo{person}{Alex Endert}.} \bibinfo{year}{2023}\natexlab{b}.
\newblock \showarticletitle{Datapilot: Utilizing quality and usage information
  for subset selection during visual data preparation}. In
  \bibinfo{booktitle}{\emph{Proceedings of the 2023 CHI Conference on Human
  Factors in Computing Systems}}. \bibinfo{pages}{1--18}.
\newblock


\bibitem[Nargesian et~al\mbox{.}(2020)]%
        {nargesian2020organizing}
\bibfield{author}{\bibinfo{person}{Fatemeh Nargesian}, \bibinfo{person}{Ken~Q.
  Pu}, \bibinfo{person}{Erkang Zhu}, \bibinfo{person}{Bahar
  Ghadiri~Bashardoost}, {and} \bibinfo{person}{Ren\'{e}e~J. Miller}.}
  \bibinfo{year}{2020}\natexlab{}.
\newblock \showarticletitle{Organizing Data Lakes for Navigation}
  \emph{(\bibinfo{series}{SIGMOD '20})}. \bibinfo{publisher}{Association for
  Computing Machinery}, \bibinfo{address}{New York, NY, USA},
  \bibinfo{pages}{1939–1950}.
\newblock
\showISBNx{9781450367356}
\href{https://doi.org/10.1145/3318464.3380605}{doi:\nolinkurl{10.1145/3318464.3380605}}


\bibitem[Nargesian et~al\mbox{.}(2019)]%
        {nargesian2019datalakemanagement}
\bibfield{author}{\bibinfo{person}{Fatemeh Nargesian}, \bibinfo{person}{Erkang
  Zhu}, \bibinfo{person}{Ren\'{e}e~J. Miller}, \bibinfo{person}{Ken~Q. Pu},
  {and} \bibinfo{person}{Patricia~C. Arocena}.}
  \bibinfo{year}{2019}\natexlab{}.
\newblock \showarticletitle{Data Lake Management: Challenges and
  Opportunities}.
\newblock \bibinfo{journal}{\emph{Proc. VLDB Endow.}} \bibinfo{volume}{12},
  \bibinfo{number}{12} (\bibinfo{date}{aug} \bibinfo{year}{2019}),
  \bibinfo{pages}{1986–1989}.
\newblock
\showISSN{2150-8097}
\href{https://doi.org/10.14778/3352063.3352116}{doi:\nolinkurl{10.14778/3352063.3352116}}


\bibitem[Paden et~al\mbox{.}(2024)]%
        {paden2024biasbuzz}
\bibfield{author}{\bibinfo{person}{Jamal~R Paden}, \bibinfo{person}{Arpit
  Narechania}, {and} \bibinfo{person}{Alex Endert}.}
  \bibinfo{year}{2024}\natexlab{}.
\newblock \showarticletitle{{BiasBuzz: Combining Visual Guidance with Haptic
  Feedback to Increase Awareness of Analytic Behavior during Visual Data
  Analysis}}. In \bibinfo{booktitle}{\emph{{Extended Abstracts of the CHI
  Conference on Human Factors in Computing Systems}}}. \bibinfo{pages}{1--7}.
\newblock
\href{https://doi.org/10.1145/3613905.3651064}{doi:\nolinkurl{10.1145/3613905.3651064}}


\bibitem[Page(2008)]%
        {page2008difference}
\bibfield{author}{\bibinfo{person}{Scott Page}.}
  \bibinfo{year}{2008}\natexlab{}.
\newblock \bibinfo{booktitle}{\emph{The difference: How the power of diversity
  creates better groups, firms, schools, and societies-new edition}}.
\newblock \bibinfo{publisher}{Princeton University Press}.
\newblock


\bibitem[P{\'e}rez-Messina et~al\mbox{.}(2022)]%
        {perez2022typology}
\bibfield{author}{\bibinfo{person}{Ignacio P{\'e}rez-Messina},
  \bibinfo{person}{Davide Ceneda}, \bibinfo{person}{Mennatallah El-Assady},
  \bibinfo{person}{Silvia Miksch}, {and} \bibinfo{person}{Fabian Sperrle}.}
  \bibinfo{year}{2022}\natexlab{}.
\newblock \showarticletitle{{A Typology of Guidance Tasks in Mixed-Initiative
  Visual Analytics Environments}}. In \bibinfo{booktitle}{\emph{{Computer
  Graphics Forum}}}, Vol.~\bibinfo{volume}{41}. Wiley Online Library,
  \bibinfo{pages}{465--476}.
\newblock


\bibitem[Ribeiro et~al\mbox{.}(2016)]%
        {ribeiro2016should}
\bibfield{author}{\bibinfo{person}{Marco~Tulio Ribeiro},
  \bibinfo{person}{Sameer Singh}, {and} \bibinfo{person}{Carlos Guestrin}.}
  \bibinfo{year}{2016}\natexlab{}.
\newblock \showarticletitle{"Why should I trust you?" Explaining the
  predictions of any classifier}. In \bibinfo{booktitle}{\emph{Proceedings of
  the 22nd ACM SIGKDD international conference on knowledge discovery and data
  mining}}. \bibinfo{pages}{1135--1144}.
\newblock


\bibitem[Richards(2006)]%
        {richards2006representational}
\bibfield{author}{\bibinfo{person}{Robert Richards}.}
  \bibinfo{year}{2006}\natexlab{}.
\newblock \showarticletitle{Representational state transfer (rest)}.
\newblock In \bibinfo{booktitle}{\emph{Pro PHP XML and web services}}.
  \bibinfo{publisher}{Springer}, \bibinfo{pages}{633--672}.
\newblock


\bibitem[Smith and Mosier(1986)]%
        {smith1986guidelines}
\bibfield{author}{\bibinfo{person}{Sidney~L Smith} {and}
  \bibinfo{person}{Jane~N Mosier}.} \bibinfo{year}{1986}\natexlab{}.
\newblock \bibinfo{booktitle}{\emph{Guidelines for designing user interface
  software}}.
\newblock \bibinfo{publisher}{Citeseer}.
\newblock


\bibitem[Sperrle et~al\mbox{.}(2022)]%
        {sperrle2022lotse}
\bibfield{author}{\bibinfo{person}{Fabian Sperrle}, \bibinfo{person}{Davide
  Ceneda}, {and} \bibinfo{person}{Mennatallah El-Assady}.}
  \bibinfo{year}{2022}\natexlab{}.
\newblock \showarticletitle{Lotse: A practical framework for guidance in visual
  analytics}.
\newblock \bibinfo{journal}{\emph{IEEE Transactions on Visualization and
  Computer Graphics}} \bibinfo{volume}{29}, \bibinfo{number}{1}
  (\bibinfo{year}{2022}), \bibinfo{pages}{1124--1134}.
\newblock


\bibitem[Sperrle et~al\mbox{.}(2024)]%
        {sperrle2024wizardofozguidance}
\bibfield{author}{\bibinfo{person}{Fabian Sperrle},
  \bibinfo{person}{Mennatallah El-Assady}, \bibinfo{person}{Alessio Arleo},
  {and} \bibinfo{person}{Davide Ceneda}.} \bibinfo{year}{2024}\natexlab{}.
\newblock \showarticletitle{A Wizard of Oz Study of Guidance Strategies and
  Dynamics}.
\newblock \bibinfo{journal}{\emph{IEEE Transactions on Visualization and
  Computer Graphics}} (\bibinfo{year}{2024}), \bibinfo{pages}{1--15}.
\newblock
\href{https://doi.org/10.1109/TVCG.2024.3418782}{doi:\nolinkurl{10.1109/TVCG.2024.3418782}}


\bibitem[Sperrle et~al\mbox{.}(2021a)]%
        {sperrle2021co}
\bibfield{author}{\bibinfo{person}{Fabian Sperrle}, \bibinfo{person}{Astrik
  Jeitler}, \bibinfo{person}{J{\"u}rgen Bernard}, \bibinfo{person}{Daniel
  Keim}, {and} \bibinfo{person}{Mennatallah El-Assady}.}
  \bibinfo{year}{2021}\natexlab{a}.
\newblock \showarticletitle{{Co-adaptive visual data analysis and guidance
  processes}}.
\newblock \bibinfo{journal}{\emph{{Computers \& Graphics}}}
  (\bibinfo{year}{2021}).
\newblock


\bibitem[Sperrle et~al\mbox{.}(2020)]%
        {sperrle2020learning}
\bibfield{author}{\bibinfo{person}{Fabian Sperrle},
  \bibinfo{person}{Astrik~Veronika Jeitler}, \bibinfo{person}{J{\"u}rgen
  Bernard}, \bibinfo{person}{Daniel~A Keim}, {and} \bibinfo{person}{Mennatallah
  El-Assady}.} \bibinfo{year}{2020}\natexlab{}.
\newblock \showarticletitle{{Learning and teaching in co-adaptive guidance for
  mixed-initiative visual analytics}}. In \bibinfo{booktitle}{\emph{{EuroVis
  Workshop on Visual Analytics (EuroVA)}}}. \bibinfo{pages}{61--65}.
\newblock


\bibitem[Sperrle et~al\mbox{.}(2021b)]%
        {sperrle2021topicmodelrefinement}
\bibfield{author}{\bibinfo{person}{Fabian Sperrle}, \bibinfo{person}{Hanna
  Sch{\"a}fer}, \bibinfo{person}{Daniel Keim}, {and}
  \bibinfo{person}{Mennatallah El-Assady}.} \bibinfo{year}{2021}\natexlab{b}.
\newblock \showarticletitle{{Learning Contextualized User Preferences for
  Co-Adaptive Guidance in Mixed-Initiative Topic Model Refinement}}. In
  \bibinfo{booktitle}{\emph{{Computer Graphics Forum}}},
  Vol.~\bibinfo{volume}{40}. Wiley Online Library, \bibinfo{pages}{215--226}.
\newblock


\bibitem[Sperrle et~al\mbox{.}(2021c)]%
        {sperrle2021learning}
\bibfield{author}{\bibinfo{person}{Fabian Sperrle}, \bibinfo{person}{Hanna
  Sch{\"a}fer}, \bibinfo{person}{Daniel Keim}, {and}
  \bibinfo{person}{Mennatallah El-Assady}.} \bibinfo{year}{2021}\natexlab{c}.
\newblock \showarticletitle{Learning Contextualized User Preferences for
  Co-Adaptive Guidance in Mixed-Initiative Topic Model Refinement}. In
  \bibinfo{booktitle}{\emph{Computer Graphics Forum}},
  Vol.~\bibinfo{volume}{40}. Wiley Online Library, \bibinfo{pages}{215--226}.
\newblock


\bibitem[Strauss and Corbin(1998)]%
        {strauss1998basics}
\bibfield{author}{\bibinfo{person}{Anselm Strauss} {and}
  \bibinfo{person}{Juliet Corbin}.} \bibinfo{year}{1998}\natexlab{}.
\newblock \bibinfo{booktitle}{\emph{{Basics of Qualitative Research: Techniques
  and Procedures for Developing Grounded Theory}}}.
\newblock \bibinfo{publisher}{{Sage Publications}}.
\newblock
\href{https://doi.org/10.4135/9781452230153}{doi:\nolinkurl{10.4135/9781452230153}}


\bibitem[Tableau(2022)]%
        {tableauprep}
\bibfield{author}{\bibinfo{person}{Tableau}.} \bibinfo{year}{2022}\natexlab{}.
\newblock \bibinfo{title}{Tableau Prep}.
\newblock
\urldef\tempurl%
\url{https://www.tableau.com/products/prep}
\showURL{%
Retrieved May 25, 2022 from \tempurl}


\bibitem[Thom-Santelli et~al\mbox{.}(2010)]%
        {thom2010you}
\bibfield{author}{\bibinfo{person}{Jennifer Thom-Santelli},
  \bibinfo{person}{Dan Cosley}, {and} \bibinfo{person}{Geri Gay}.}
  \bibinfo{year}{2010}\natexlab{}.
\newblock \showarticletitle{What do you know? Experts, novices and
  territoriality in collaborative systems}. In
  \bibinfo{booktitle}{\emph{Proceedings of the SIGCHI Conference on Human
  Factors in Computing Systems}}. \bibinfo{pages}{1685--1694}.
\newblock


\bibitem[Tynan and Drayton(1987)]%
        {tynan1987market}
\bibfield{author}{\bibinfo{person}{A~Caroline Tynan} {and}
  \bibinfo{person}{Jennifer Drayton}.} \bibinfo{year}{1987}\natexlab{}.
\newblock \showarticletitle{Market segmentation}.
\newblock \bibinfo{journal}{\emph{Journal of marketing management}}
  \bibinfo{volume}{2}, \bibinfo{number}{3} (\bibinfo{year}{1987}),
  \bibinfo{pages}{301--335}.
\newblock


\bibitem[Tzini and Jain(2018)]%
        {tzini2018role}
\bibfield{author}{\bibinfo{person}{Konstantina Tzini} {and}
  \bibinfo{person}{Kriti Jain}.} \bibinfo{year}{2018}\natexlab{}.
\newblock \showarticletitle{The role of anticipated regret in advice taking}.
\newblock \bibinfo{journal}{\emph{Journal of Behavioral Decision Making}}
  \bibinfo{volume}{31}, \bibinfo{number}{1} (\bibinfo{year}{2018}),
  \bibinfo{pages}{74--86}.
\newblock


\bibitem[Vedadi et~al\mbox{.}(2021)]%
        {vedadi2021herd}
\bibfield{author}{\bibinfo{person}{Ali Vedadi}, \bibinfo{person}{Merrill
  Warkentin}, {and} \bibinfo{person}{Alan Dennis}.}
  \bibinfo{year}{2021}\natexlab{}.
\newblock \showarticletitle{Herd behavior in information security
  decision-making}.
\newblock \bibinfo{journal}{\emph{Information \& Management}}
  \bibinfo{volume}{58}, \bibinfo{number}{8} (\bibinfo{year}{2021}),
  \bibinfo{pages}{103526}.
\newblock


\bibitem[Vodrahalli et~al\mbox{.}(2022)]%
        {vodrahalli2022humans}
\bibfield{author}{\bibinfo{person}{Kailas Vodrahalli}, \bibinfo{person}{Roxana
  Daneshjou}, \bibinfo{person}{Tobias Gerstenberg}, {and}
  \bibinfo{person}{James Zou}.} \bibinfo{year}{2022}\natexlab{}.
\newblock \showarticletitle{Do humans trust advice more if it comes from ai? an
  analysis of human-ai interactions}. In \bibinfo{booktitle}{\emph{Proceedings
  of the 2022 AAAI/ACM Conference on AI, Ethics, and Society}}.
  \bibinfo{pages}{763--777}.
\newblock


\bibitem[Wall et~al\mbox{.}(2022)]%
        {wall2021lrg}
\bibfield{author}{\bibinfo{person}{Emily Wall}, \bibinfo{person}{Arpit
  Narechania}, \bibinfo{person}{Adam Coscia}, \bibinfo{person}{Jamal Paden},
  {and} \bibinfo{person}{Alex Endert}.} \bibinfo{year}{2022}\natexlab{}.
\newblock \showarticletitle{{Left, Right, and Gender: Exploring Interaction
  Traces to Mitigate Human Biases}}.
\newblock \bibinfo{journal}{\emph{{IEEE Transactions on Visualization and
  Computer Graphics}}} \bibinfo{volume}{28}, \bibinfo{number}{1}
  (\bibinfo{year}{2022}), \bibinfo{pages}{966--975}.
\newblock
\href{https://doi.org/10.1109/TVCG.2021.3114862}{doi:\nolinkurl{10.1109/TVCG.2021.3114862}}


\bibitem[Weber and Johnson(2009)]%
        {weber2009mindful}
\bibfield{author}{\bibinfo{person}{Elke~U Weber} {and} \bibinfo{person}{Eric~J
  Johnson}.} \bibinfo{year}{2009}\natexlab{}.
\newblock \showarticletitle{Mindful judgment and decision making}.
\newblock \bibinfo{journal}{\emph{Annual review of psychology}}
  \bibinfo{volume}{60}, \bibinfo{number}{1} (\bibinfo{year}{2009}),
  \bibinfo{pages}{53--85}.
\newblock


\bibitem[Yaniv(2004)]%
        {yaniv2004receiving}
\bibfield{author}{\bibinfo{person}{Ilan Yaniv}.}
  \bibinfo{year}{2004}\natexlab{}.
\newblock \showarticletitle{Receiving other people’s advice: Influence and
  benefit}.
\newblock \bibinfo{journal}{\emph{Organizational behavior and human decision
  processes}} \bibinfo{volume}{93}, \bibinfo{number}{1} (\bibinfo{year}{2004}),
  \bibinfo{pages}{1--13}.
\newblock


\bibitem[Zeelenberg et~al\mbox{.}(1998)]%
        {zeelenberg1998experience}
\bibfield{author}{\bibinfo{person}{Marcel Zeelenberg}, \bibinfo{person}{Wilco~W
  Van~Dijk}, \bibinfo{person}{Antony SR~Manstead}, {and}
  \bibinfo{person}{Joopvan der Pligt}.} \bibinfo{year}{1998}\natexlab{}.
\newblock \showarticletitle{The experience of regret and disappointment}.
\newblock \bibinfo{journal}{\emph{Cognition \& Emotion}} \bibinfo{volume}{12},
  \bibinfo{number}{2} (\bibinfo{year}{1998}), \bibinfo{pages}{221--230}.
\newblock


\bibitem[Zhang et~al\mbox{.}(2003)]%
        {dataprep2003zhang}
\bibfield{author}{\bibinfo{person}{Shichao Zhang}, \bibinfo{person}{Chengqi
  Zhang}, {and} \bibinfo{person}{Qiang Yang}.} \bibinfo{year}{2003}\natexlab{}.
\newblock \showarticletitle{Data preparation for data mining}.
\newblock \bibinfo{journal}{\emph{Applied Artificial Intelligence}}
  \bibinfo{volume}{17}, \bibinfo{number}{5-6} (\bibinfo{year}{2003}),
  \bibinfo{pages}{375--381}.
\newblock
\href{https://doi.org/10.1080/713827180}{doi:\nolinkurl{10.1080/713827180}}
\showeprint{https://doi.org/10.1080/713827180}


\bibitem[Zhang and Ives(2020)]%
        {zhang2020finding}
\bibfield{author}{\bibinfo{person}{Yi Zhang} {and} \bibinfo{person}{Zachary~G.
  Ives}.} \bibinfo{year}{2020}\natexlab{}.
\newblock \showarticletitle{{Finding Related Tables in Data Lakes for
  Interactive Data Science}}. In \bibinfo{booktitle}{\emph{Proceedings of the
  2020 ACM SIGMOD International Conference on Management of Data}}.
  \bibinfo{publisher}{Association for Computing Machinery},
  \bibinfo{address}{New York, NY, USA}, \bibinfo{pages}{1951–1966}.
\newblock
\showISBNx{9781450367356}


\end{thebibliography}

\appendix

\section{Summary of Hypotheses \textbf{H1--H3}}
\label{section:appendix-h1h2h3}


\begin{table*}
    \centering
    \renewcommand{\arraystretch}{0.75} 
    \begin{tabular}{r|l|c|r}
    
    & \textbf{Summary of Hypotheses Testing: H1--H3} & \textbf{Validation} & \textbf{\emph{p}-value} \\
    \addlinespace\hline\addlinespace

    \textbf{H1} & \textbf{Hypothesis}: \guidance{} participants who receive guidance will \emph{be less uncertain} & & \\

    & \emph{about their attribute selections} than \control{} participants who do not receive guidance. & & \\

    & & & \\

    & \textbf{Metric}: Variance of \# interactions across all attributes. &  & \\

    & \textbf{Result}: \guidance{} (median=19.19), \control{} (median=21.82) & & \\

    & \hspace{0.85cm}  \includegraphics[width=0.30\linewidth]{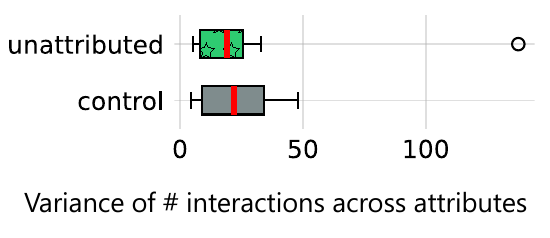} & & \\

    & \textbf{Statistical Test}: \guidance{} < \control{} & \faCheckCircle & 0.34 \\

    \addlinespace\hline\addlinespace

    \textbf{H2} & \textbf{Hypothesis}: \guidance{} participants who receive guidance will \emph{select more attributes} & &  \\

    & than \control{} participants who do not receive guidance. & & \\

    & & & \\

    & \textbf{Metric}: Total \# attributes selected at the end of the task. & & \\

    & \textbf{Result}: \guidance{} (median=10.5), \control{} (median=6) & & \\

    & \hspace{0.85cm}  \includegraphics[width=0.30\linewidth]{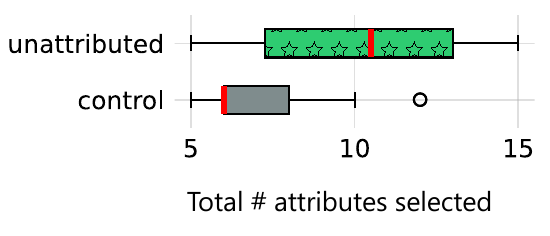} & & \\

     & \textbf{Statistical Test}: \guidance{} > \control{} & \aptLtoX{\includegraphics[width=6pt]{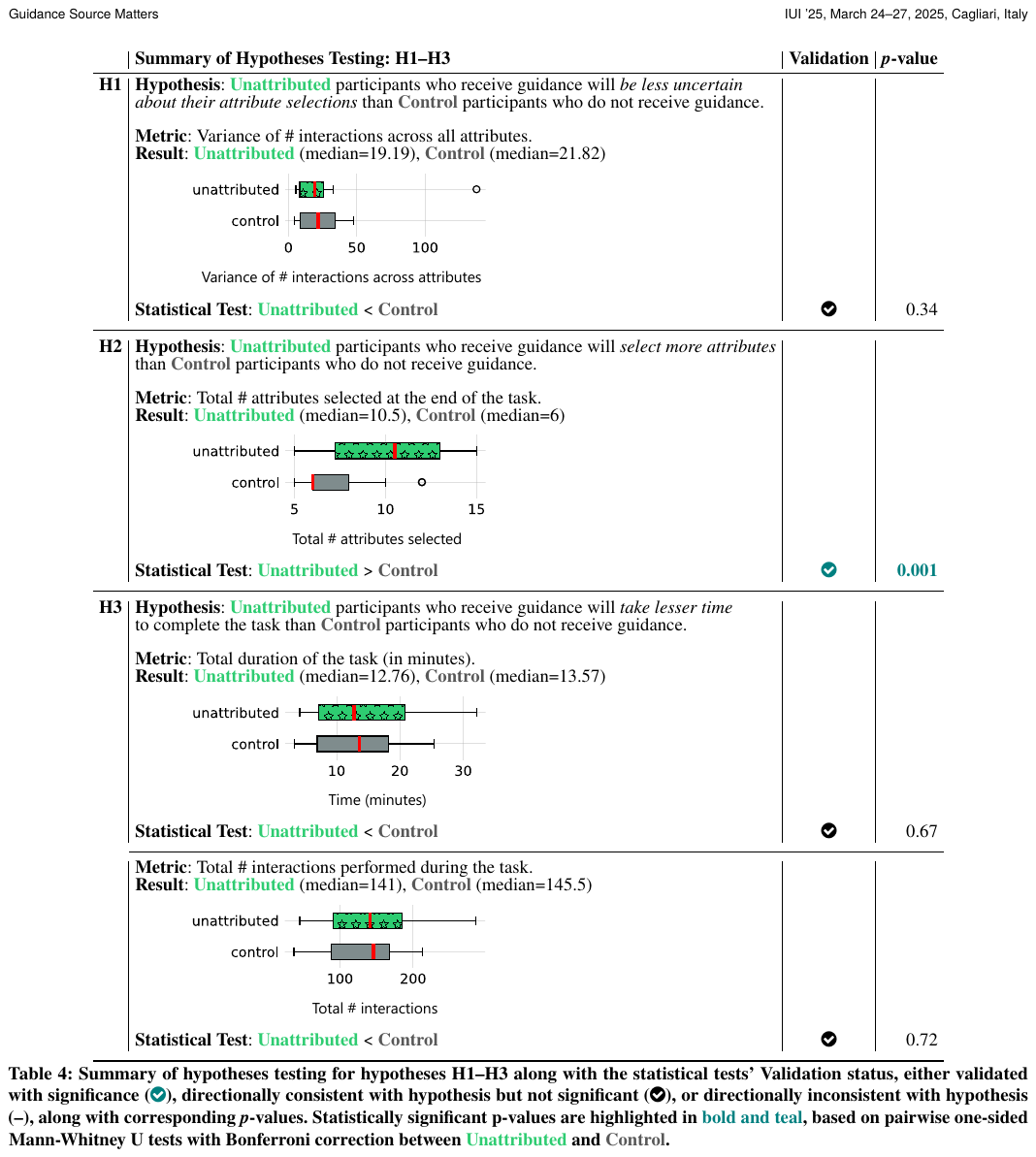}}{\textcolor{teal}{\faCheckCircle}} & \textcolor{teal}{\textbf{0.001}} \\

    \addlinespace\hline\addlinespace

    \textbf{H3} & \textbf{Hypothesis}: \guidance{} participants who receive guidance will \emph{take lesser time} &  \\

    & to complete the task than \control{} participants who do not receive guidance. & & \\

    & & & \\

    & \textbf{Metric}: Total duration of the task (in minutes). &  &  \\

    & \textbf{Result}: \guidance{} (median=12.76), \control{} (median=13.57) & & \\

    & \hspace{0.85cm}  \includegraphics[width=0.30\linewidth]{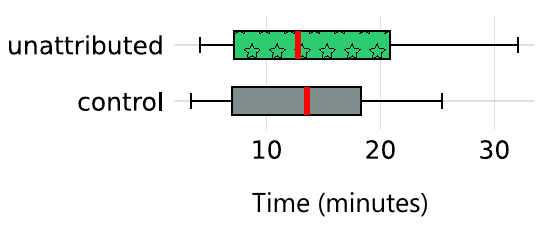} & & \\

    & \textbf{Statistical Test}: \guidance{} < \control{} & \faCheckCircle & 0.67 \\

    \addlinespace\cline{2-4}\addlinespace

    & \textbf{Metric}: Total \# interactions performed during the task. & & \\

    & \textbf{Result}: \guidance{} (median=141), \control{} (median=145.5) & & \\

    & \hspace{0.85cm}  \includegraphics[width=0.30\linewidth]{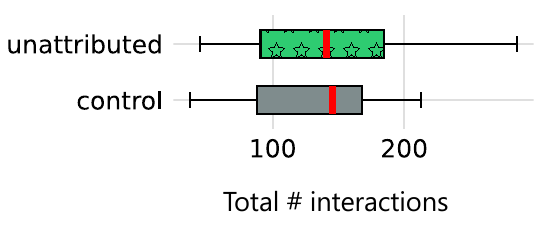} & & \\

    & \textbf{Statistical Test}: \guidance{} < \control{} & \faCheckCircle & 0.72 \\

    \addlinespace\bottomrule
    
    \end{tabular}
    \caption{Summary of hypotheses testing for hypotheses \textbf{H1}--\textbf{H3} along with the statistical tests' \textbf{Validation} status, either validated with significance (\includegraphics[width=8pt]{tealfacirc.pdf}), directionally consistent with hypothesis but not significant (\faCheckCircle), or directionally inconsistent with hypothesis (--), along with corresponding \textbf{\emph{p}-values}. Statistically significant p-values are highlighted in \textcolor{teal}{\textbf{bold and teal}}, based on pairwise one-sided Mann-Whitney U tests with Bonferroni correction between \guidance{} and \control{}.}
    \label{table:results-hypotheses-summary-h123}
\end{table*}


\section{Summary of Hypotheses \textbf{H4--H5}}
\label{section:appendix-h4h5}


\begin{table*}
    \centering
    \renewcommand{\arraystretch}{0.70} 
    \begin{tabular}{r|l|c|r}
    
    & \textbf{Summary of Hypotheses Testing: H4--H5} & \textbf{Validation} & \textbf{\emph{p}-value} \\
    \addlinespace\hline\addlinespace

    \textbf{H4} & \textbf{Hypothesis}: Participants will find guidance to \emph{have more utility} when it comes from \expert{} > \ai{} > \group{} &  &  \\

    & & & \\

    & \textbf{Metric}: Total \# attributes selected at the end of the task. & & \\

    & \textbf{Result}: \expert{} (median=9), \ai{} (median=9), \group{} (median=7) & & \\

    & \hspace{0.85cm}  \includegraphics[width=0.30\linewidth]{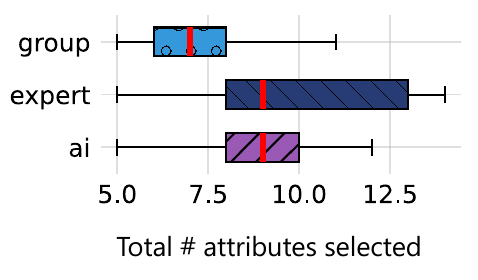} & & \\

    & \textbf{Statistical Test}: & & \\
    
    & \hspace{1.05cm} \expert{} > \ai{} & -- & 1.0 \\
    & \hspace{1.05cm} \expert{} > \group{} & \aptLtoX{\includegraphics[width=6pt]{tealfacirc.pdf}}{\textcolor{teal}{\faCheckCircle}} & \textcolor{teal}{\textbf{0.01}} \\
    & \hspace{1.05cm} \ai{} > \group{} & \aptLtoX{\includegraphics[width=6pt]{tealfacirc.pdf}}{\textcolor{teal}{\faCheckCircle}} & \textcolor{teal}{\textbf{0.04}} \\

    \addlinespace\hline\addlinespace
    
    \textbf{H5} & \textbf{Hypothesis}: Participants will \emph{verify the guidance more} when it comes from \ai{} > \expert{} > \group{}. &  & \\

    & & & \\

    & \textbf{Metric}: Difference in \# interactions with guided attributes after and before guidance. & & \\

    & \textbf{Result}: \ai{} (median=31), \expert{} (median=14), \group{} (median=12.5) &  \\

    & \hspace{0.85cm} \includegraphics[width=0.30\linewidth]{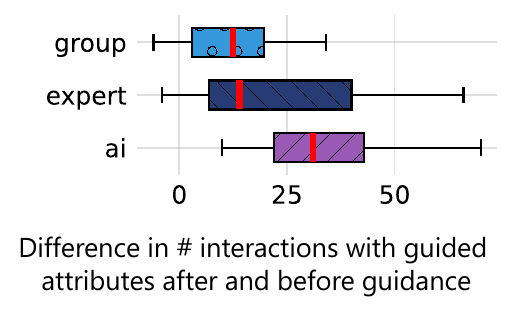} & & \\
    
    & \textbf{Statistical Test}: & & \\
    
    & \hspace{1.05cm} \ai{} > \expert{} & \faCheckCircle & 0.11 \\
    & \hspace{1.05cm} \ai{} > \group{} & \aptLtoX{\includegraphics[width=6pt]{tealfacirc.pdf}}{\textcolor{teal}{\faCheckCircle}} & \textcolor{teal}{\textbf{0.003}} \\
    & \hspace{1.05cm} \expert{} > \group{} & \faCheckCircle & 0.59 \\

    \addlinespace\cline{2-4}\addlinespace
        
    & \textbf{Metric}: Ratio of \# interactions with guided attributes to \# interactions with all attributes. &  &  \\

    & \textbf{Result}: \ai{} (median=0.36), \expert{} (median=0.23), \group{} (median=0.19) & & \\
    
    & \hspace{0.85cm}  \includegraphics[width=0.34\linewidth]{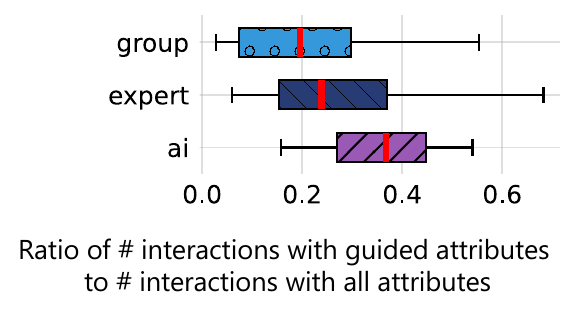} & & \\

    & \textbf{Statistical Test}: & & \\
    
    & \hspace{1.05cm} \ai{} > \expert{} & \faCheckCircle & 0.27 \\
    & \hspace{1.05cm} \ai{} > \group{} & \aptLtoX{\includegraphics[width=6pt]{tealfacirc.pdf}}{\textcolor{teal}{\faCheckCircle}} & \textcolor{teal}{\textbf{0.01}} \\
    & \hspace{1.05cm} \expert{} > \group{} & \faCheckCircle & 0.35 \\
 
    \addlinespace\bottomrule
    
    \end{tabular}
    \caption{Summary of hypotheses testing for hypotheses \textbf{H4}--\textbf{H5} along with the statistical tests' \textbf{Validation} status, either validated with significance (\includegraphics[width=6pt]{tealfacirc.pdf}), directionally consistent with hypothesis but not significant (\faCheckCircle), or directionally inconsistent with hypothesis (--), along with corresponding \textbf{\emph{p}-values}. Statistically significant p-values are highlighted in \textcolor{teal}{\textbf{bold and teal}}, based on pairwise one-sided Mann-Whitney U tests with Bonferroni correction between \ai{}, \expert{}, and \group{}.}
    \label{table:results-hypotheses-summary-h45}
\end{table*}

\section{Summary Statistics of the Pre-, and Post- Study Questionnaires}
\label{section:appendix-questionnaires}

\begin{table*}
    \centering
    \caption{Median scores about questions asked during the two pre-study and one post-study questionnaires, on a scale from 1 (Low/Disagree/None at all) to 7 (High/Agree/A lot). \underline{\textbf{Bold and underline}} indicate the largest values in each row.}
    \label{tab:questionnaire-scores}
    \renewcommand{\arraystretch}{1.1} 
    \begin{tabular}{lc|c|c|c|c}
        \toprule
        \multirow{2}{*}{\textbf{Metric}} & \multicolumn{1}{c|}{\textbf{\ai{}}} & \multicolumn{1}{c|}{\textbf{\expert{}}} & \multicolumn{1}{c|}{\textbf{\group{}}} & \multicolumn{1}{c|}{\textbf{\guidance{}}} & \multicolumn{1}{c}{\textbf{\control{}}} \\
        \cline{2-6}
        & \color{black}{Median} & \color{black}{Median} & \color{black}{Median} & \color{black}{Median} & \color{black}{Median} \\
        \hline
        \multicolumn{6}{c}{\cellcolor{lightergray}Pre Study Questionnaire 1: Questions about Guidance. Asked to all participants. Scale: 1 (Disagree) - 7 (Agree).} \\
        \hline
        Guidance Can Be \textbf{Beneficial} & \textcolor{colorai}{\underline{\textbf{6.0}}} & \textcolor{colorexpert}{\underline{\textbf{6.0}}} & \textcolor{colorgroup}{\underline{\textbf{6.0}}} & \textcolor{colorguidance}{\underline{\textbf{6.0}}} & \underline{\textbf{6.0}} \\
        \hline
        \multicolumn{6}{c}{\cellcolor{lightergray}Pre Study Questionnaire 2: Questions about Guidance. Not asked to \control{}. Scale: 1 (Disagree) - 7 (Agree).} \\
        \hline
        Guidance Can Be \textbf{Reliable} & \textcolor{colorai}{4.0} & \textcolor{colorexpert}{5.0} & \textcolor{colorgroup}{5.0} & \textcolor{colorguidance}{\underline{\textbf{6.0}}} & - \\
        Guidance Can Increase \textbf{Confidence} & \textcolor{colorai}{4.0} & \textcolor{colorexpert}{\underline{\textbf{6.0}}} & \textcolor{colorgroup}{4.0} & \textcolor{colorguidance}{\underline{\textbf{6.0}}} & - \\
        Guidance Can Add \textbf{Value} & \textcolor{colorai}{5.0} & \textcolor{colorexpert}{5.0} & \textcolor{colorgroup}{5.0} & \textcolor{colorguidance}{\underline{\textbf{6.0}}} & - \\
        \hline
        \multicolumn{6}{c}{\cellcolor{lightergray}Post Study Questionnaire: Questions on Guidance Received during Task. Not asked to \control{}. Scale: 1 (Disagree) - 7 (Agree).} \\
        \hline
        Guidance Was \textbf{Trustworthy} & \textcolor{colorai}{3.0} & \textcolor{colorexpert}{4.0} & \textcolor{colorgroup}{3.0} & \textcolor{colorguidance}{\textbf{5.0}} & - \\
        Guidance Came From A \textbf{Knowledgeable} Source & \underline{\textbf{5.0}} & \textcolor{colorexpert}{4.0} & \textcolor{colorgroup}{\underline{\textbf{5.0}}} & \textcolor{colorguidance}{4.0} & - \\
        Guidance Was \textbf{Reliable} & \textcolor{colorai}{3.0} & \textcolor{colorexpert}{4.0} & \textcolor{colorgroup}{3.0} & \textcolor{colorguidance}{\underline{\textbf{4.5}}} & - \\
        Guidance Increased \textbf{Confidence} & \textcolor{colorai}{3.0} & \textcolor{colorexpert}{3.0} & \textcolor{colorgroup}{3.0} & \textcolor{colorguidance}{\underline{\textbf{4.0}}} & - \\
        Guidance Was \textbf{Relevant} & \textcolor{colorai}{4.0} & \textcolor{colorexpert}{4.0} & \textcolor{colorgroup}{3.0} & \textcolor{colorguidance}{\underline{\textbf{5.0}}} & - \\
        Guidance Helped Avoid Potential \textbf{Pitfalls} & \textcolor{colorai}{2.0} & \textcolor{colorexpert}{\underline{\textbf{3.0}}} & \textcolor{colorgroup}{\underline{\textbf{3.0}}} & \textcolor{colorguidance}{\underline{\textbf{3.0}}} & - \\
        Guidance Gave Valuable \textbf{Suggestions} & \textcolor{colorai}{\underline{\textbf{5.0}}} & \textcolor{colorexpert}{3.0} & \textcolor{colorgroup}{3.0} & \textcolor{colorguidance}{\underline{\textbf{5.0}}} & - \\
        Guidance \textbf{Added Value} & \textcolor{colorai}{\underline{\textbf{5.0}}} & \textcolor{colorexpert}{3.0} & \textcolor{colorgroup}{3.0} & \textcolor{colorguidance}{3.5} & - \\
        Guidance Was \textbf{Appropriate} & \textcolor{colorai}{4.0} & \textcolor{colorexpert}{4.0} & \textcolor{colorgroup}{4.0} & \textcolor{colorguidance}{\underline{\textbf{5.5}}} & - \\
        \hline
        \multicolumn{6}{c}{\cellcolor{lightergray}Post Study Questionnaire: Questions on Guidance Received during Task. Not asked to \control{}. Scale: 1 (None at all) - 7 (A lot).} \\
        \hline
        How Much Did You \textbf{Rely} on Guidance & \textcolor{colorai}{\underline{\textbf{3.0}}} & \textcolor{colorexpert}{2.0} & \textcolor{colorgroup}{2.0} & \textcolor{colorguidance}{\underline{\textbf{3.0}}} & - \\
        How Much Did You \textbf{Regret} Relying on Guidance & \textcolor{colorai}{\underline{\textbf{5.0}}} & \textcolor{colorexpert}{2.0} & \textcolor{colorgroup}{2.0} & \textcolor{colorguidance}{3.5} & - \\
        \hline
        \multicolumn{6}{c}{\cellcolor{lightergray}Post Study Questionnaire: Questions about Guidance. Asked only to \control{}. Scale: 1 (Disagree) - 7 (Agree).} \\
        \hline
        Guidance Can Be \textbf{Trustworthy} & \textcolor{colorai}{-} & \textcolor{colorexpert}{-} & \textcolor{colorgroup}{-} & \textcolor{colorguidance}{-} & \underline{\textbf{5.0}} \\
        Guidance Can Show It Came From A \textbf{Knowledgeable} Source & \textcolor{colorai}{-} & \textcolor{colorexpert}{-} & \textcolor{colorgroup}{-} & \textcolor{colorguidance}{-} & \underline{\textbf{6.0}} \\
        Guidance Can Be \textbf{Reliable} & \textcolor{colorai}{-} & \textcolor{colorexpert}{-} & \textcolor{colorgroup}{-} & \textcolor{colorguidance}{-} & \underline{\textbf{5.0}} \\
        Guidance Can Increase \textbf{Confidence} & \textcolor{colorai}{-} & \textcolor{colorexpert}{-} & \textcolor{colorgroup}{-} & \textcolor{colorguidance}{-} & \underline{\textbf{7.0}} \\
        Guidance Can Be \textbf{Relevant} & \textcolor{colorai}{-} & \textcolor{colorexpert}{-} & \textcolor{colorgroup}{-} & \textcolor{colorguidance}{-} & \underline{\textbf{6.0}} \\
        Guidance Can Help Avoid Potential \textbf{Pitfalls} & \textcolor{colorai}{-} & \textcolor{colorexpert}{-} & \textcolor{colorgroup}{-} & \textcolor{colorguidance}{-} & \underline{\textbf{6.0}} \\
        Guidance Can Give Valuable \textbf{Suggestions} & \textcolor{colorai}{-} & \textcolor{colorexpert}{-} & \textcolor{colorgroup}{-} & \textcolor{colorguidance}{-} & \underline{\textbf{6.5}} \\
        Guidance Can Add \textbf{Value} & \textcolor{colorai}{-} & \textcolor{colorexpert}{-} & \textcolor{colorgroup}{-} & \textcolor{colorguidance}{-} & \underline{\textbf{6.0}} \\
        Guidance Can Be \textbf{Appropriate} & \textcolor{colorai}{-} & \textcolor{colorexpert}{-} & \textcolor{colorgroup}{-} & \textcolor{colorguidance}{-} & \underline{\textbf{6.0}} \\
        \hline
        \multicolumn{6}{c}{\cellcolor{lightergray}Post Study Questionnaire: (What-If) Questions about Guidance. Condition-specific. Scale: 1 (Low) - 7 (High).} \\
        \hline
        How Much \textbf{Faith} Would You Have If Guidance From \ai{} &\textcolor{colorai}{-} & \textcolor{colorexpert}{\underline{\textbf{4.0}}} & \textcolor{colorgroup}{\underline{\textbf{4.0}}} & \textcolor{colorguidance}{3.0} & 3.5 \\
        How Much \textbf{Faith} Would You Have If Guidance From \expert{} & \textcolor{colorai}{6.0} & \textcolor{colorexpert}{-} & \textcolor{colorgroup}{5.0} & \textcolor{colorguidance}{6.0} & \underline{\textbf{6.5}} \\
        How Much \textbf{Faith} Would You Have If Guidance From \group{} & \textcolor{colorai}{\underline{\textbf{6.0}}} & \textcolor{colorexpert}{\underline{\textbf{6.0}}} & \textcolor{colorgroup}{-} & \textcolor{colorguidance}{\underline{\textbf{6.0}}} & \underline{\textbf{6.0}} \\
        \hline
        How Much Would You \textbf{Prefer} Guidance If It Comes From \ai{} & \textcolor{colorai}{-} & \textcolor{colorexpert}{-} & \textcolor{colorgroup}{-} & \textcolor{colorguidance}{\underline{\textbf{5.0}}} & \underline{\textbf{5.0}} \\
        How Much Would You \textbf{Prefer} Guidance If It Comes From \expert{} & \textcolor{colorai}{-} & \textcolor{colorexpert}{-} & \textcolor{colorgroup}{-} & \textcolor{colorguidance}{6.0} & \underline{\textbf{6.5}} \\
        How Much Would You \textbf{Prefer} Guidance If It Comes From \group{} & \textcolor{colorai}{-} & \textcolor{colorexpert}{-} & \textcolor{colorgroup}{-} & \textcolor{colorguidance}{\underline{\textbf{6.0}}} & \underline{\textbf{6.0}} \\
        How Much Would You \textbf{Rely} on Guidance If It Comes From \ai{} & \textcolor{colorai}{-} & \textcolor{colorexpert}{-} & \textcolor{colorgroup}{-} & \textcolor{colorguidance}{4.0} & \underline{\textbf{4.5}} \\
        How Much Would You \textbf{Rely} on Guidance If It Comes From \expert{} & \textcolor{colorai}{-} & \textcolor{colorexpert}{-} & \textcolor{colorgroup}{-} & \textcolor{colorguidance}{\underline{\textbf{6.0}}} & \underline{\textbf{6.0}} \\
        How Much Would You \textbf{Rely} on Guidance If It Comes From \group{} & \textcolor{colorai}{-} & \textcolor{colorexpert}{-} & \textcolor{colorgroup}{-} & \textcolor{colorguidance}{\underline{\textbf{6.0}}} & \underline{\textbf{6.0}} \\
        \hline
        \multicolumn{6}{c}{\cellcolor{lightergray}Post Study Questionnaire: Questions about Prior Experiences with Guidance. Scale: 1 (Low) - 7 (High).} \\
        \hline
        How Often Have You \textbf{Sought} Guidance In Life & \textcolor{colorai}{-} & \textcolor{colorexpert}{-} & \textcolor{colorgroup}{-} & \textcolor{colorguidance}{\underline{\textbf{5.0}}} & - \\
        How Often Have You \textbf{Sought} Guidance In Life From \ai{} & \textcolor{colorai}{3.0} & \textcolor{colorexpert}{4.0} & \textcolor{colorgroup}{4.0} & \textcolor{colorguidance}{4.0} & \underline{\textbf{5.0}} \\
        How Often Have You \textbf{Sought} Guidance In Life From \expert{} & \textcolor{colorai}{\underline{\textbf{5.0}}} & \textcolor{colorexpert}{\underline{\textbf{5.0}}} & \textcolor{colorgroup}{4.0} & \textcolor{colorguidance}{3.5} & 4.0 \\
        How Often Have You \textbf{Sought} Guidance In Life From \group{} & \textcolor{colorai}{3.0} & \textcolor{colorexpert}{\underline{\textbf{5.0}}} & \textcolor{colorgroup}{2.0} & \textcolor{colorguidance}{3.0} & 4.0 \\
        \hline
    \end{tabular}
\end{table*}

\end{document}